\documentclass[iop]{emulateapj}
\usepackage{color}
\shorttitle{The RINGS Survey III: \Ha\ \FP\ Dataset}
\shortauthors{Mitchell et al.}
\definecolor{red}{rgb}{0.7,0.1,0.1}

\long\def\Ignore#1{{\relax}}

\slugcomment{Revised version submitted to AJ on January 23, 2018}

\def\lcdm{$\rm{\Lambda}$CDM}
\def\h{$^\textrm{h}$}
\def\m{$^\textrm{m}$}
\def\FP{Fabry-P\'erot}
\def\Ha{H$\alpha$}
\def\N2{[\ion{N}{2}]}
\def\df{\texttt{DiskFit}}

\begin{document}

\title{The RINGS Survey III: Medium-Resolution H$\alpha$ Fabry-P\'erot
  Kinematic Dataset}

\author{Carl J. Mitchell} \affiliation{Department of Physics and
  Astronomy, Rutgers, the State University of New Jersey, 136
  Frelinghuysen Rd., Piscataway, NJ 08854}

\author{J. A. Sellwood\altaffilmark{1}} \affiliation{Department of
  Physics and Astronomy, Rutgers, the State University of New Jersey,
  136 Frelinghuysen Rd., Piscataway, NJ 08854} \affil{$^1$ Current
  address: Steward Observatory, 933 N Cherry Ave, Tucson, AZ 85721}

\author{T. B. Williams} \affiliation{South African Astronomical
  Observatory, Observatory, Cape Town 7925, South Africa\\}
\affiliation{Astronomy Department, University of Cape Town, Rondebosch
  7700, South Africa\\} \affiliation{Department of Physics and
  Astronomy, Rutgers, the State University of New Jersey, 136
  Frelinghuysen Rd., Piscataway, NJ 08854}

\author{Kristine Spekkens} \affiliation{Department of Physics, Royal
  Military College of Canada, P.O. Box 17000, Station Forces,
  Kingston, Ontario K7K 7B4, Canada}

\author{Rachel Kuzio de Naray} \affiliation{Department of Physics and
  Astronomy, Georgia State University, Atlanta, GA 30302}

\author{Alex Bixel} \affiliation{Department of Astronomy and Steward
  Observatory, The University of Arizona, 933 North Cherry Ave,
  Tucson, AZ 85721}

\email{cmitchell@physics.rutgers.edu}
\email{sellwood@as.arizona.edu}
\email{williams@saao.ac.za}
\email{kristine.spekkens@rmc.ca}
\email{kuzio@astro.gsu.edu}
\email{abixel@email.arizona.edu}

\date{\today}

\begin{abstract}
The distributions of stars, gas, and dark matter in disk galaxies
provide important constraints on galaxy formation models, particularly
on small spatial scales ($<$1~kpc).  We have designed the RSS Imaging
spectroscopy Nearby Galaxy Survey (RINGS) to target a sample of 19
nearby spiral galaxies.  For each of these galaxies, we are obtaining
and modeling \Ha\ and \ion{H}{1}~21~cm spectroscopic data as well as
multi-band photometric data.  We intend to use these models to explore
the underlying structure and evolution of these galaxies in a
cosmological context, as well as whether the predictions of \lcdm\ are
consistent with the mass distributions of these galaxies.  In this
paper, we present spectroscopic imaging data for 14 of the RINGS
galaxies observed with the medium spectral resolution \FP\ etalon on
the Southern African Large Telescope.  From these observations, we
derive high spatial resolution line of sight velocity fields of the
\Ha\ line of excited hydrogen, as well as maps and azimuthally
averaged profiles of the integrated \Ha\ and \N2\ emission and oxygen
abundances.  We then model these kinematic maps with axisymmetric
models, from which we extract rotation curves and projection
geometries for these galaxies.  We show that our derived rotation
curves agree well with other determinations and the similarity of the
projection angles with those derived from our photometric images
argues against these galaxies having intrinsically oval disks.
\end{abstract}

\keywords{galaxies, galaxies: spiral, galaxies: individual, galaxies:
  kinematics and dynamics}

\section{Introduction}
\label{sec:intro}
The standard cosmological paradigm of Cold Dark Matter with the
addition of a cosmological constant (\lcdm) has been successful at
interpreting astrophysical phenomena on a wide range of scales, from
the large scale structure of the Universe to the formation of
individual galaxies \citep{2015ARA&A..53...51S}.  However, it remains
somewhat unclear whether the internal structures of simulated galaxies
formed in a \lcdm\ framework are consistent with observations of real
galaxies.

In spiral galaxies, the structure of dark matter halos can be
constrained using galaxy rotation curves
\citep[e.g.][]{1978PhDT.......195B}.  Typically, the observed rotation
curve is decomposed into contributions from stars and gas and any
remaining velocity is attributed to dark matter.  In cosmological
simulations of dark matter structure growth, dark matter halos have
been observed to follow a broken power law form
\citep[e.g.][]{1965TrAlm...5...87E, 1996ApJ...462..563N,
  2004MNRAS.349.1039N, 2008MNRAS.387..536G}.  To account for the
additional gravitational pull provided by baryons, modifications can
be applied to theoretical halo density profiles to increase their
densities at small radii \citep[e.g.][]{2004ApJ...616...16G,
  2005ApJ...634...70S}.  Applying these modified halo models to
observed rotation curves produces dark matter halos which are
underdense relative to the predictions of \lcdm\ simulations
\citep{2015A&A...574A.113P}.

Numerical simulations which incorporate stellar feedback in galaxies
have partially eased this tension by showing that feedback from
baryonic processes can redistribute dark matter within a galaxy
\citep{2010Natur.463..203G, 2012MNRAS.421.3464P,
  2013MNRAS.429.3068T}.  These effects are stronger in galaxies with
lower masses \citep[e.g.][]{2011AJ....142...24O, 2011MNRAS.415.1051B,
  2014Natur.506..171P}.  Recent simulations have shown that the ability
of a galaxy to redistribute dark matter through stellar feedback
depends on the ratio of its stellar mass to its halo mass
\citep[e.g.][]{2014MNRAS.437..415D, 2015MNRAS.454.1719B}.  These
$M_*/M_{\rm halo}$-dependent density profiles have been shown by
\citet{2017MNRAS.466.1648K} to be more consistent with the photometry
and rotation curves of real galaxies than traditional NFW profiles.

The relationship between dark matter halos and observed rotation
curves is not a trivial one, as measurements of rotation curves can be
biased by non-circular motions, projection effects, and halo
triaxiality \citep[e.g.][]{2004ApJ...617.1059R, 2006MNRAS.373.1117H,
  2007ApJ...657..773V}.  Measurements of one-dimensional rotation
curves are therefore insufficient to constrain the three-dimensional
mass distributions.  All of these mechanisms for potential bias in
rotation curves leave kinematic signatures in the full
three-dimensional velocity distributions of galaxy disks.  For example,
gas streaming along bars and spiral arms has both circular and radial
components to its velocity, and therefore will affect the line of
sight velocities along the major and minor axes differently
\citep{2010MNRAS.404.1733S}.

Measurements of the velocity field of the entire disk at high spatial
resolution are required to extract these kinematic signatures.  For
example, to separate bar-like flows in spiral galaxies from their
rotation curves, $<200~$pc spatial resolution is required
\citep[e.g.][]{2007ApJ...659.1176M, 2010MNRAS.404.1733S,
  2014A&A...568A..70B,2015MNRAS.451.4397H}.

In recent years, the state of the art in numerical simulations has
moved to smaller and smaller spatial scales.  However, comparisons of
these simulations to observed galaxies have been lacking, partially
due to a lack of velocity fields of sufficiently high resolution for
comparison.

We have designed the RSS Imaging spectroscopy Nearby Galaxy Survey
(RINGS) to obtain the high-resolution kinematic data necessary to
probe these open questions of galaxy structure.  Our survey targets 19
nearby, late-type spiral galaxies over a wide range of masses (67 km
s$^{-1} < V_{\rm flat} < $ 275 km s$^{-1}$) and luminosities (-17.5 $> M_V
>$ -21.5).  The survey is designed to exploit the large collecting area
and large field-of-view of the Robert Stobie Spectrograph (RSS) on the
Southern African Large Telescope (SALT).  In addition to the high
spatial resolution \Ha\ kinematic data from SALT's RSS, we are
obtaining lower spatial resolution \ion{H}{1} 21~cm kinematic
observations and have obtained $BVRI$ photometric imaging of these
galaxies.

A number of previous surveys have obtained two-dimensional
\Ha\ velocity fields of galaxies with similar goals to RINGS, e.g.
BH$\alpha$BAR \citep{2005MNRAS.360.1201H}, GHASP
\citep{2008MNRAS.388..500E}, GH$\alpha$FaS
\citep{2008PASP..120..665H}, DiskMass \citep{2010ApJ...716..198B}, and
CALIFA \citep{2012A&A...538A...8S}.  Compared to these surveys, our
data are deeper and more extended thanks to SALT's large primary
mirror and large angular field-of-view.  The typical angular
resolution of the RINGS data is similar to that of the DiskMass and
CALIFA surveys and somewhat worse than that of GH$\alpha$FaS.
However, the RINGS galaxies are typically more nearby than the
galaxies in those surveys, and our physical resolutions are comparable
to those of GH$\alpha$FaS and higher than those of DiskMass and
CALIFA.  The typical spectral resolution of our data ($R\sim1300$) is
similar to that of CALIFA ($R\sim1000$) and lower than that of
DiskMass ($R\sim8000$) and GH$\alpha$FaS ($R\sim15000$).  Our target
selection criteria also differ from these surveys in choosing a
representative sample of partially inclined galaxies across a wide
range of Hubble classifications, masses, and luminosities.

In Paper I \citep{RINGS1}, we presented our first \Ha\ and \ion{H}{1}
kinematic data and modelling for the galaxy NGC 2280.  In Paper II
\citep{RINGSPhot}, we presented our photometric sample and modelling.
In this paper, we present kinematic maps and axisymmetric models of 14
of the 19 RINGS galaxies.  The maps are derived from data taken using
the medium-resolution etalon of SALT's \FP\ system.  The typical
angular resolution of our resulting \Ha\ velocity fields is
$\sim2.5\arcsec$, corresponding to a typical spatial resolution of
$\sim250~$pc at the source locations.  We then model the kinematic
data using the \df\ software package \citep{2007ApJ...664..204S,
  2010MNRAS.404.1733S} and show that the derived rotation curves
generally agree well with others in the literature.  We also compare
the fitted projection parameters with those obtained from our {\it I}-band
images.  Finally, we present azimuthally-averaged \Ha\ and \N2
profiles for these galaxies, which we use to derive oxygen abundance
gradients.  In future papers in this series, we will use our velocity
maps in order to better understand these galaxies' mass distributions.

\begin{deluxetable*}{llcccccccccc}
\tablewidth{0pt}
\tablecaption{RINGS Medium-Resolution \FP Observations\label{tab:tab1}}
\tablehead{
\colhead{Galaxy} &
\colhead{Class} &
\colhead{Obs. Date} &
\colhead{Exp. Time} &
\colhead{Seeing} &
\colhead{$\sigma_{\lambda}$ [\AA]} &
\colhead{N$_{pix}$} &
\colhead{N$_{elem}$} &
\colhead{$D$ [Mpc]} &
\colhead{Scale [pc/\arcsec]} &
\colhead{Seeing} &
\colhead{$M_I$}}
\startdata
NGC 337A & SAB(s)dm & 11 Sept 2012 & 22$\times$100s & 1.7\arcsec & 0.054 & 9448 & 1842 & 2.57\tablenotemark{a} & 12.5 & 30~pc & -16.7\\
 &  & 10 Oct 2012 & 26$\times$91s & 2.1\arcsec & 0.028 &  &  &  &  &  & \\
 &  & 12 Oct 2012 & 26$\times$38s & \textbf{2.4}\arcsec & 0.064 &  &  &  &  &  & \\
\hline
NGC 578 & SAB(rs)c & 29 Dec 2011 & 28$\times$50s & 1.8\arcsec & 0.025 & 29890 & 4416 & 27.1\tablenotemark{b} & 131 & 370~pc & -22.5\\
 &  & 23 Oct 2012 & 23$\times$98s & \textbf{2.8}\arcsec & 0.036 &  &  &  &  &  & \\
\hline
NGC 908 & SA(s)c & 1 Nov 2011 & 41$\times$60s & 2.2\arcsec & 0.033 & 28045 & 4284 & 19.4\tablenotemark{b} & 94.1 & 220~pc & -21.6\\
 &  & 28 Dec 2011 & 25$\times$100s & \textbf{2.3}\arcsec & 0.034 &  &  &  &  &  & \\
\hline
NGC 1325 & SA(s)bc & 1 Nov 2011 & 24$\times$90s & 2.0\arcsec & 0.025 & 7813 & 1532 & 23.7\tablenotemark{b} & 115 & 310~pc & -21.3\\
 &  & 28 Dec 2011 & 23$\times$100s & \textbf{2.7}\arcsec & 0.026 &  &  &  &  &  & \\
\hline
NGC 1964 & SAB(s)b & 2 Apr 2012 & 23$\times$70s & 2.4\arcsec & 0.025 & 12093 & 2220 & 20.9\tablenotemark{b} & 101 & 270~pc & -21.8\\
 &  & 1 Feb 2013 & 25$\times$80s & \textbf{2.7}\arcsec & 0.031 &  &  &  &  &  & \\
\hline
NGC 2280 & SA(s)cd & 1 Nov 2011 & 25$\times$60s & 2.0\arcsec & 0.040 & 27198 & 6609 & 24.0\tablenotemark{b} & 116 & 260~pc & -20.8\\
 &  & 28 Dec 2011 & 26$\times$50s & \textbf{2.2}\arcsec & 0.049 &  &  &  &  &  & \\
\hline
NGC 3705 & SAB(r)ab & 1 Feb 2013 & 23$\times$77s & 2.3\arcsec & 0.10 & 6687 & 1394 & 18.5\tablenotemark{b} & 89.7 & 230~pc & -19.9\\
 &  & 26 Feb 2014 & 23$\times$80s & \textbf{2.6}\arcsec & 0.064 &  &  &  &  &  & \\
\hline
NGC 4517A & SB(rs)dm & 23 Apr 2012 & 37$\times$80s & 2.5\arcsec & 0.049 & 2904 & 592 & 26.7\tablenotemark{b} & 129 & 360~pc & -22.8\\
 &  & 27 Apr 2015 & 20$\times$90s & \textbf{2.8}\arcsec & 0.070 &  &  &  &  &  & \\
 &  & 27 Apr 2015 & 21$\times$102s & 2.3\arcsec & 0.19 &  &  &  &  &  & \\
 &  & 7 May 2015 & 21$\times$95s & 2.3\arcsec & 0.21 &  &  &  &  &  & \\
 &  & 7 May 2015 & 21$\times$100s & 1.9\arcsec & 0.21 &  &  &  &  &  & \\
\hline
NGC 4939 & SA(s)bc & 14 Apr 2013 & 24$\times$90s & 1.9\arcsec & 0.051 & 18971 & 4809 & 41.6\tablenotemark{b} & 202 & 420~pc & -22.9\\
 &  & 27 Apr 2015 & 24$\times$95s & \textbf{2.1}\arcsec & 0.062 &  &  &  &  &  & \\
\hline
NGC 5364 & SA(rs)bc pec & 28 May 2012 & 24$\times$80s & \textbf{2.0}\arcsec & 0.087 & 14756 & 4720 & 18.1\tablenotemark{c} & 87.8 & 180~pc & -21.2\\
\hline
NGC 6118 & SA(s)cd & 28 May 2012 & 22$\times$100s & \textbf{2.0}\arcsec & 0.052 & 14207 & 3686 & 22.9\tablenotemark{b} & 111 & 220~pc & -22.7\\
 &  & 2 Sept 2012 & 22$\times$85s & 1.8\arcsec & 0.038 &  &  &  &  &  & \\
\hline
NGC 6384 & SAB(r)bc & 15 July 2014 & 23$\times$85s & 2.5\arcsec & 0.062 & 17442 & 3760 & 19.7\tablenotemark{d} & 95.5 & 260~pc & -21.8\\
 &  & 31 July 2014 & 23$\times$85s & \textbf{2.7}\arcsec & 0.077 &  &  &  &  &  & \\
\hline
NGC 7606 & SA(s)b & 17 Aug 2014 & 13$\times$87s & 2.2\arcsec & 0.073 & 10454 & 1835 & 34.0\tablenotemark{e} & 165 & 460~pc & -24.4\\
 &  & 1 Sept 2014 & 26$\times$85s & 1.7\arcsec & 0.11 &  &  &  &  &  & \\
 &  & 6 Aug 2015 & 22$\times$92s & \textbf{2.8}\arcsec & 0.15 &  &  &  &  &  & \\
\hline
NGC 7793 & SA(s)d & 2 Sept 2014 & 22$\times$90s & 2.1\arcsec & 0.027 & 101908 & 12028 & 3.44\tablenotemark{f} & 16.7 & 50~pc & -18.5\\
 &  & 3 Sept 2014 & 18$\times$90s & 2.7\arcsec & 0.072 &  &  &  &  &  & \\
 &  & 8 June 2015 & 22$\times$90s & 2.6\arcsec & 0.10 &  &  &  &  &  & \\
 &  & 14 Aug 2015 & 18$\times$90s & 3.0\arcsec & 0.13 &  &  &  &  &  & \\
 &  & 21 Aug 2015 & 20$\times$80s & \textbf{3.0}\arcsec & 0.084 &  &  &  &  &  & \\
\enddata
\tablecomments{A summary of our observations and resulting kinematic maps for the 14 galaxies presented here. From left to right, columns are: (1) galaxy name, (2) morphological classification, (3) observation date, (4) number of exposures and time per exposure, (5) effective seeing with worst seeing for each galaxy marked in bold, (6) estimated uncertainty in our wavelength solution\tablenotemark{g}, (7) number of pixels in our fitted maps, (8) number of independent resolution elements in our fitted maps, (9) redshift-independent distance and reference, (10) angular scale at the distances in column 9, (11) seeing in physical units at the distances in column 9, and (12) absolute \textit{I}-band magnitude derived from the photometry of \citet{RINGSPhot} and the distances in column 9.
\tablenotetext{a}{\citet{1985AAS...59...43B} using \textit{B}-band isophotal diameter Tully-Fisher relation.}
\tablenotetext{b}{\citet{1997ApJS..109..333W} using \textit{H}-band Tully-Fisher relation.}
\tablenotetext{c}{\citet{2007AA...465...71T} using \textit{H}-band Tully-Fisher relation.}
\tablenotetext{d}{\citet{2000ApJ...540..634P} using SN Ia \textit{B}- and \textit{V}-band light curves (SN 1971L).}
\tablenotetext{e}{\citet{1997ApJS..109..333W} using \textit{I}-band Tully-Fisher relation.}
\tablenotetext{f}{\citet{2010AJ....140.1475P} using 17 Cepheid variable stars.}
\tablenotetext{g}{At the wavelength of \Ha, a wavelength shift of 0.1 \AA\ corresponds to a velocity shift of 4.6~km~s$^{-1}$}}
\end{deluxetable*}

\section{Data Acquisition and Reduction}
\label{sec:datareduction}
We obtained data on 14 nearby late-type galaxies with the
medium-resolution mode of the \FP\ interferometer on the RSS of SALT.
Our data were acquired over a total exposure time of 19 hours during
the period 11 Nov 2011 to 8 Sept 2015.  A typical single observation
consists of $\sim25$ exposures, each of length $\sim70~$seconds.  The
medium-resolution etalon has a spectral full width at half maximum
(FWHM) at \Ha\ of $\sim5~$\AA.  For each exposure taken in an
observation, we offset the wavelength of the etalon's peak
transmission by $\sim2~$\AA\ from the previous exposure.  Each
observation therefore represents a scan over a $\sim50~$\AA\ range in
$\sim2~$\AA\ steps.  For each galaxy, we attempted to obtain at least
two such observations.  A summary of the properties of these 14
galaxies and our observations is provided in Table \ref{tab:tab1}.

Note that NGC 2280, which we have discussed previously in
\citet{RINGS1}, is among the galaxies presented in this work.  Because
several aspects of our data reduction process have changed somewhat
(e.g.\ flat-field correction and ghost subtraction, discussed below)
since that work was published, we have chosen to present an updated
velocity field of that galaxy here to ensure homogeneity across the
final sample.

\subsection{Preliminary Data Reduction}
We have utilized the PySALT\footnote{http://pysalt.salt.ac.za/}
\citep{pysalt} software package to perform preliminary reductions of
our raw SALT images.  The tasks in PySALT apply standard routines for
gain variation corrections, bias subtraction, CCD crosstalk
corrections, and cosmic ray removal.

\subsection{Flattening}
\label{sec:flattening}
The unusual design of SALT introduces unique challenges in calibrating
the intensity of our images.  SALT's primary
mirror\footnote{https://www.salt.ac.za/telescope/\#telescope-primary-mirror}
is composed of a hexagonal grid of 91 1-meter mirrors.  Unlike most
telescopes, the primary mirror remains stationary over the course of
an observation and object tracking is accomplished by moving the
secondary optics package in the primary mirror's focal plane.  The
full collecting area of the primary mirror is rarely utilized, as some
mirror segments are unable to illuminate the secondary depending on a
target's position.  Overall, the available collecting area of the
primary mirror is smaller by $\sim30$\% at the beginning and end of an
observation relative to the middle.

The individual mirror segments are removed for realuminization and
replaced on $\sim$weekly timescales in a sequential scheme.  This
results in the reflectivity of the primary mirror varying as a
function of position on the mirror, and these variations change over
time as different mirror segments are freshly realuminized.

As a target galaxy passes through SALT's field of view, individual
mirror segments pass in and out of the secondary payload's field of
view, changing the fraction of the total collecting area utilized as a
function of time.

Furthermore, differential vignetting of images occurs within the
spherical aberration corrector (SAC) on the secondary payload.  This
effect also varies as a function of object position overhead (as the
secondary package moves through the focal plane to track an
object).  This vignetting effect changes image intensities by
$\sim5-10$\% across an image.

The combined effects of these factors result in image intensity
variations which are: position-dependent within a single image,
pointing-dependent over the course of an observation as the target
drifts overhead, and time-dependent over the $\sim$weekly
segment-replacement timescale.

A traditional approach to flat-field calibration (i.e.\ combining
several exposures of the twilight sky) is insufficient for correcting
these effects, as this approach will not account for the
pointing-dependent effects.  Theoretical modelling of the sensitivity
variations by ray-tracing software is not feasible due to the frequent
replacement of mirror segments with different reflective properties.

In a previous paper \citep{RINGS1}, we utilized an approach for NGC
2280 which compared stellar photometry in our SALT \FP\ images to
\textit{R}-band images from the CTIO 0.9m telescope
\citep{RINGSPhot}.  For $\sim50$ stars present in both sets of images,
we computed an intensity ratio between our SALT images and the
\textit{R}-band image.  For each SALT image, we then fitted a quadratic
two-dimensional polynomial to these intensity ratios.  By scaling each
of our images by its corresponding polynomial, we were able to correct
for these variations.

Unlike NGC 2280, most of our target galaxies do not overlap with dense
star fields and we therefore cannot apply this approach.  Instead, we
have developed a new approach which utilizes the night sky background
to calibrate our photometry.  We make the assumption that the intrinsic
night sky background has uniform intensity over the $8\arcmin$ field
of view over the course of each individual exposure ($\sim70~$s).  We
then mask objects in our fields using a sigma-clipped cutoff for stars
and a large elliptical mask for the galaxy.  We fit the remaining
pixels with a quadratic two-dimensional polynomial of the same form
used in the stellar photometry approach described above.  We then scale
the pixel values in each image by this fitted polynomial.  If the
assumption of uniform sky brightness is valid, this method results in
a uniformly illuminated field.

In order to validate the assumption of uniform sky intensity, we have
applied this ``sky-fitting'' approach to our data on NGC 2280 and
compared it to our previous ``star-fitting'' approach for the same
data.  We found no significant differences in the resulting fitted
polynomials for either of the two nights for which we had data on that
galaxy.  This suggests that the sky-fitting approach is sufficient for
flattening our images.  The assumption of a uniform sky background is
less likely to be valid if a target galaxy fills a large fraction of
the field of view, as is the case with our observations of NGC
7793.  We have examined several spectra obtained from overlapping
observations of this galaxy, and it appears any errors introduced by a
non-uniform sky background are small compared to other sources of
uncertainty.

We utilize this ``sky-fitting'' approach to flat-field correction for
all 14 of the galaxies presented in this work.

\begin{figure*}
  \begin{center}
    \includegraphics[width=\hsize]{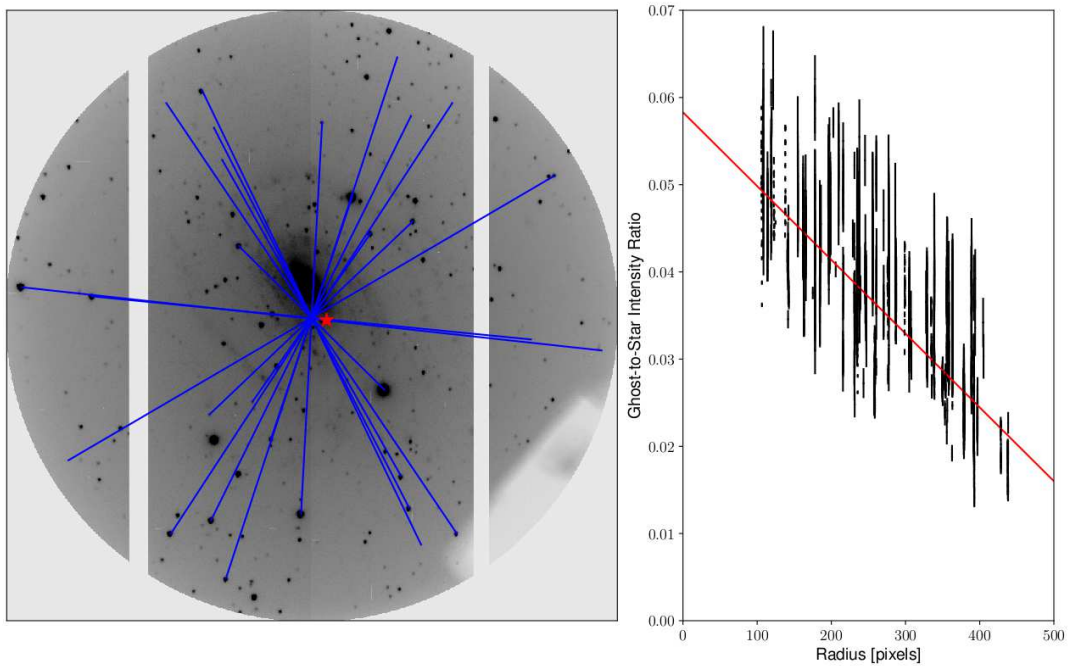}
  \end{center}
  \caption{Left: A median-combined image of our 15 July 2014
    observations of NGC 6384 with detected star-ghost pairs marked
    with blue lines.  The large red star marks the location of the
    point about which the intensity ratios of a ghost to its star are
    symmetric.  The large rectangular feature in the lower-right
    portion of the left panel is the shadow of SALT's tracking probe
    and the affected pixels have been masked from any calculations.
    Right: The black points with error bars mark the intensity ratios
    between ghosts and stars as a function of radius from the point
    marked in the left panel.  These star/ghost pairs were selected
    from all of our SALT \FP\ observations.  The solid red line shows
    our linear fit to these intensity ratios.
    \label{fig:ghosts}}
\end{figure*}

\subsection{Ghost identification and subtraction}
\label{sec:ghosts}
Reflections between the \FP\ etalon and the CCD detector result in
each light source in an image appearing twice -- once at its true
position and again at a reflected position, known as the ``diametric
ghost'' \citep{ghosts}.  The positions of these reflections are
symmetric about a single point in the image, the location at which the
instrument's optical axis intersects the plane of the CCD.  The left
panel of Figure \ref{fig:ghosts} illustrates this effect in one of our
observations of NGC 6384.

As will be discussed in \S \ref{sec:wavecal}, the wavelength
calibration solutions for our images are symmetric about the same
central point.  The ghost positions are therefore extremely useful for
precisely determining the location of this point.  By matching each
star in an image to its ghost and averaging their positions, we are
able to determine our reflection centers to within a small fraction of
a pixel.

While useful for determining the location of the symmetry axis, the
presence of these ghosts adversely affects our goal of measuring
velocities.  In particular, the reflected image of a target galaxy
often overlaps with the galaxy itself.  This effect is extremely
undesirable, since it mixes emission from gas at one location and
velocity with emission from gas at a different location and velocity.

In order to remove them, we perform aperture photometry on each
star-ghost pair to determine intensity ratios between the ghosts and
their real counterparts.  These ratios are typically $\sim5$\%.  In a
previous paper \citep{RINGS1}, we simply rotated each image by
180\arcdeg\ about its symmetry axis and subtracted a small multiple of
the rotated image from the original.  After examining a much larger
quantity of data, it appears that the intensity ratio between an
object and its ghost depends linearly on the object's distance from a
central point.  This decreasing ghost intensity ratio is caused by
vignetting within the camera optics of the non-telecentric reflection
from the CCD.  This central point's location is not coincident with
the center of reflection (private communication: D. O'Donoghue), but
appears to be consistent among all of our observations.  The right
panel of Figure \ref{fig:ghosts} shows the dependence of the ghost
intensity ratio on radius from this point.  We have fitted a linear
function to the flux ratios of star-ghost pairs in several of our
observations, which decreases from $\sim6$\% at the central point to
$\sim2$\% at the edge of the images.  We then apply the same
reflect-and-subtract approach as in \citep{RINGS1}, except that here
we rescale the reflected images by this linear function rather than a
constant factor.  This process removes most of the ghost image
intensity from our science images without necessitating masking of
these regions.

\vfill\eject
\subsection{Alignment and Normalization}
\label{sec:align_norm}
Among the images of a single observation, we use the centroid
locations of several stars to align our images to one another.
Typically, the image coordinate system drifts by
$\sim0.25$\arcsec\ over the course of an observation.

As mentioned previously, different fractions of SALT's primary mirror
are utilized over the course of a single observation.  Thus, the
photometric sensitivity of each image varies over an observational
sequence.  To correct for this effect, we perform aperture photometry
on the same stars which were used for aligning the images in order to
determine a normalization factor for each image.  We then scale each
image by a multiplicative normalization factor so that each of these
stars has the same intensity in all of our images.  Typically, between
10 and 50 stars are used in this process, though in some extreme cases
(e.g.\ NGC 578), the number of stars in the images can be as low as 5.

The combined effects of flattening uncertainty
(\S\ref{sec:flattening}), ghost subtraction (\S\ref{sec:ghosts}), and
normalization uncertainty (\S\ref{sec:align_norm}) result in a typical
photometric uncertainty of $\sim 10-12\%$.

When combining multiple observations which were taken at different
telescope pointings, we have utilized the \texttt{astrometry.net}
software package \citep{astrometry} to register our images' pixel
positions to accurate sky coordinates.  We then use the resulting
astrometric solutions to align our observations to one another.

Just as we used stellar photometry to normalize images from among a
single observation sequence, we use the same photometry to normalize
different observation sequences to one another.  Stars which are
visible in only one pointing are not useful for this task, so we use
the photometry of stars which are visible in more than one observation
sequence.

\subsection{Wavelength Calibration}
\label{sec:wavecal}
Collimated light incident on the \FP\ etalon arrives at different
angles depending on position in our images.  Different angles of
incidence result in different wavelengths of constructive
interference.  Thus, the peak wavelength of an image varies across the
image itself.  The wavelength of peak transmission is given by
\begin{equation}
    \lambda_{\rm peak}(R) = \frac{\lambda_{\rm cen}}{(1+R^2/F^2)^{1/2}}
\end{equation}
where $\lambda_{\rm cen}$ is the peak wavelength at the center of the
image, $R$ is the radius of a pixel from the image center, and $F$ is
the effective focal length of the camera optics, measured in units of
pixels.  The image center is the location where the optical axis
intersects the image plane, and is notably the same as the center of
the star-ghost reflections discussed in \S\ref{sec:ghosts}.

The peak wavelength at the center is determined by a parameter, $z$,
which controls the spacing of the etalon's parallel plates.  It may
also be a function of time, as a slight temporal drift in the etalon
spacing has been observed.  In general, we find that the function
\begin{equation}
  \lambda_{\rm cen}(z,t) = A + Bz + Et
  \label{eqn:wavesoln}
\end{equation}
is sufficient to describe the central wavelength's dependence on the
control parameter and time.  This equation equivalent to the one found
by \citet{rangwala} with the addition of a term which is linear in
time to account for a slight temporal drift.  We find that their
higher-order terms proportional to $z^2$ and $z^3$ are not necessary
over our relatively narrow wavelength range.

Across a single image, the wavelength of peak transmission depends
only on the radius, $R$.  Therefore, a monochromatic source which
uniformly illuminates the field will be imaged as a symmetric ring
around the image center, with radius $R_{\rm ring} =
F(\lambda_{\rm cen}^2/\lambda_{\rm ring}^2-1)^{1/2}$.

Before and after each observation sequence, exposures of neon lamps
were taken for the purposes of wavelength calibrations, which create
bright rings in the images.  Additionally, several atmospheric
emission lines of hydrogen, \N2, and OH are imaged as dim rings in our
observations of the RINGS galaxies.  By measuring the radii of these
rings, we can determine best-fitting values for the constants $A$,
$B$, $E$, and $F$ in the above equations using a least-squares
minimization fit.  We then use these fitted parameters to calibrate
the wavelengths in our images.  The sixth column of Table
\ref{tab:tab1} shows the uncertainty in each observation's wavelength
solution, calculated as the root mean square residual to our
wavelength solution divided by the square root of the number of
degrees of freedom in the fit.

\subsection{Sky Subtraction}
\label{sec:skysub}
The sky background radiation in our images is composed of two
components: a continuum, which we treat as constant with wavelength,
and emission lines from molecules in the atmosphere.

Once a wavelength solution has been found for our images, we search in
our images for ring signatures of known atmospheric emission lines
\citep{osterbrock}.  We fit for such emission lines and subtract the
fitted profiles from our images.  Occasionally, additional emission
lines are seen (as prominent rings) even after such subtraction.
These emission lines fall into two broad categories: adjacent spectral
orders and diffuse interstellar bands.

The medium-resolution \FP\ system has a free spectral range (FSR) at
\Ha\ of $\sim75~$\AA.  Thus, an atmospheric emission line $\pm75~$\AA\
from an image's true wavelength may appear in the image due to the
non-zero transmission of the order-blocking filter at
$\pm75~$\AA.  Several such emission lines have been detected in our
data and subsequently fitted and subtracted from our images.

In several of our observations, we have detected emission consistent
with the diffuse interstellar band (DIB) wavelength at 6613~\AA\
\citep{dibs}.  DIBs are commonly seen as absorption lines in stellar
spectra, and are not often observed in emission \citep{herbig95}.  This
emission has also been fitted and subtracted from our data in the same
fashion as the known night-sky emission lines.  The DIB emission was
detected in our observations of NGC 908, NGC 1325, and NGC 2280.

Once ring features from emission lines have been fitted and
subtracted, we have run a sigma-clipped statistics algorithm to
determine the typical value of the night sky continuum emission.  This
continuum value is then subtracted from each of our images before we
produce our final data cube.

\subsection{Convolution to Uniform Seeing}
Because atmospheric turbulence and mirror alignment do not remain
constant over the course of an observation, each of our images has a
slightly different value for the effective seeing FWHM.  In producing a
data cube, we artificially smear all of our images to the seeing of
the worst image of the observation track.  In principle, we could
choose to keep only images with better effective seeing and discard
images with worse seeing.  When our observations were obtained, SALT
did not have closed-loop control of the alignment of the primary
mirror segments.  Thus the image quality tended to degrade over an
observational sequence.  Discarding poorer images would therefore tend
to preferentially eliminate the longer wavelength images, since we
usually stepped upward in wavelength over the sequence.  Discarding
images would also reduce the overall depth of our observations.  For
these reasons, we choose to not discard any images when producing the
final data cubes presented in this work.

The correction to uniform seeing is done by convolution with a
Gaussian beam kernel with $\sigma_{\rm beam}^2 = \sigma_{\rm worst}^2 -
\sigma_{\rm image}^2$.  We also shift the position of the convolution
kernel's center by the values of the shifts calculated from stellar
centroids described in \S\ref{sec:align_norm}.  In this way, we
shift and convolve our images simultaneously.  The ``Seeing'' column of
Table \ref{tab:tab1} lists the worst seeing FWHM from each of our
observations.  Typical worst seeing values are between 2\arcsec\ and
3\arcsec.  In the cases where we combine multiple observations of the
same object, we convolve all observations to the seeing of the worst
image from among all observations of that object, then combine the
results into a single data cube.

\begin{figure*}
  \begin{center}
    \includegraphics[width=\hsize]{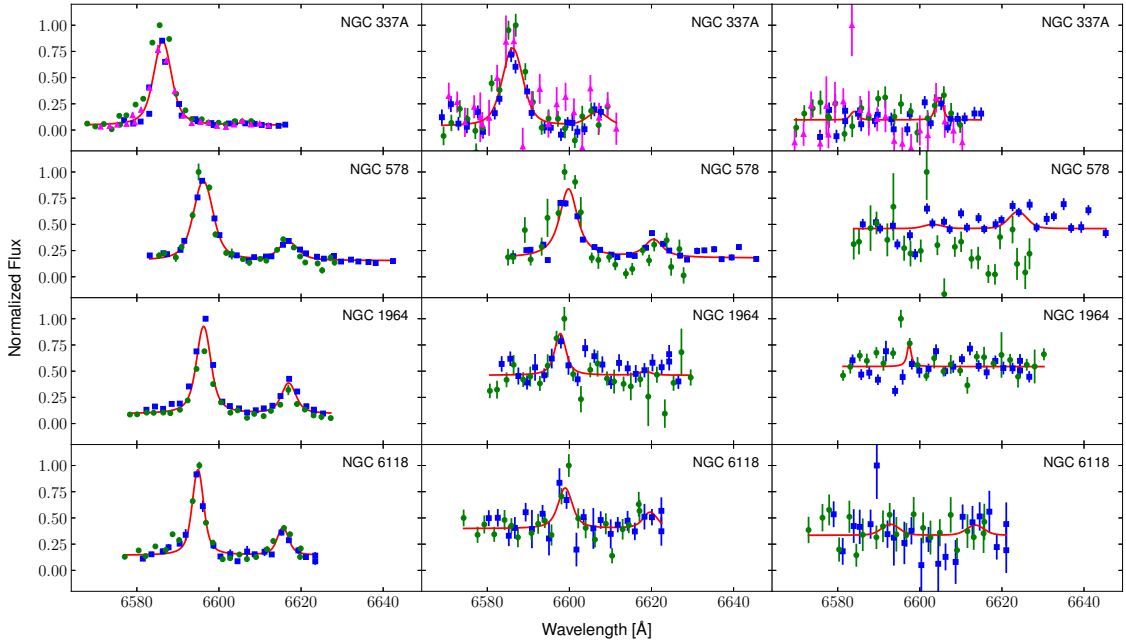}
  \end{center}
  \caption{Selected spectra (solid points with error bars) and
    best-fitting line profiles (solid red lines) from our data cubes.
    The left panels show pixels with very high signal-to-noise.  The
    middle panels show pixels with much lower signal-to-noise.  The
    right panels show pixels very low signal-to-noise which are just
    above our detection thresholds.  All spectra have been normalized
    so that the maximum value of each spectrum is 1.  Each row's
    spectra are different pixels selected from a single galaxy's data
    cube.  The different colors and shapes of points correspond to
    observations from different nights.
    \label{fig:lineprofs}}
\end{figure*}

\subsection{Line Profile Fitting}
\label{sec:linefits}
In addition to observing the \Ha\ line, our wavelength range is wide
enough to detect the \N2 6583 line as well.  We fit for both of these
lines in our spectra simultaneously.  The transmission profile of the
\FP\ etalon is well-described by a Voigt function,
\begin{equation}
    V(\lambda;\sigma_g,\gamma_l) = \int_{-\infty}^{\infty}
    G(\lambda',\sigma_g)\Gamma(\lambda-\lambda',\gamma_l)d\lambda',
\end{equation}
where $G(\lambda,\sigma_g)$ and $\Gamma(\lambda,\gamma_l)$ are
Gaussian and Lorentzian functions, respectively.  Calculating this
convolution of functions is computationally expensive, and we
therefore make use of the pseudo-Voigt function described by
\citet{voigt}.  At each spatial pixel in our data cubes, we fit a
6-parameter model of the form
\begin{eqnarray}
I(\lambda; C, F_H, F_N, \lambda_H, \sigma_g, \gamma_l) = 
\phantom{junkjunk} \nonumber \\ 
\phantom{junkjunk} C + F_HV(\lambda-\lambda_H;\sigma_g,\gamma_l) + \nonumber \\
\phantom{junkjunk} F_NV(\lambda-1.003137\lambda_H;\sigma_g,\gamma_l),
\label{eqn:model}
\end{eqnarray}
where $I(\lambda;\ldots)$ is the image intensity as a function of
wavelength and the 6 model parameters are: $C$, the continuum surface
brightness, $F_H$, the integrated surface brightness of the \Ha\ line,
$F_N$, the integrated surface brightness of the \N2 6583 line,
$\lambda_H$, the peak wavelength of Doppler-shifted \Ha, and
$\sigma_g$ and $\gamma_l$ the two line widths of the Voigt profile.
We assume that the \Ha\ and \N2 6583 emission arise from gas at the
same velocity, and the factor of 1.003137 in the above equation
reflects this assumption.

An anonymous referee questioned whether $C$ would really be constant
over the fitted range because the stellar continuum would have an
\Ha\ absorption feature at almost the same wavelength as the
\Ha\ emission we are attempting to measure.  While there may be some
effect of stellar \Ha\ absorption on the emission line strength, it is
unlikely to exactly cancel the gaseous emission, and would leave a
distorted spectral profile (e.g.\ with emission core and absorption
wings), which we do not see.  \citet{2002MNRAS.332..283R} find that
stellar absorption in disk galaxies has the greatest effect at
H$\delta$ and H$\epsilon$, and essentially no contribution at \Ha.
This suggests that absorption has a minimal effect on our estimate of
\Ha\ line strength.  Since there is no significant absorption of the
\N2 lines, we do not expect stellar absorption lines to reduce our
ability to detect emission from excited gas to any significant extent.
Estimates of the \Ha/\N2 line intensity ratio would be affected by any
\Ha\ absorption and, if important, would compromise {\em all}
spectroscopic estimates of this line intensity ratio, not exclusively
those from \FP\ data.

\begin{deluxetable*}{lcccccccc}
\tablewidth{0pt}
\tablecaption{Best-Fitting Axisymmetric \df Model Parameters\label{tab:tab2}}
\tablehead{
\colhead{Galaxy} &
\colhead{Dist [Mpc]} &
\colhead{Scale [pc/\arcsec]} &
\colhead{RA$_{\rm cen}$ [J2000]} &
\colhead{Dec$_{\rm cen}$ [J2000]} &
\colhead{$V_{\rm sys}$ [km~s$^{-1}$]} &
\colhead{$i$ [\arcdeg]} &
\colhead{PA [\arcdeg]} &
\colhead{$\chi^2$/d.o.f.}}
\startdata
NGC 337A & 2.57 & 12.5 & 01\h01\m32\fs3 $\pm$ 0\fs29 & -07\arcdeg35\arcmin23.9\arcsec $\pm$ 2.0\arcsec & 1074.3 $\pm$ 2.2 & 56.6 $\pm$ 3.4 & 77.8 $\pm$ 10.5 & 1.2 \\
NGC 578 & 27.1 & 131 & fixed\phantom{\h} & fixed\phantom{\h} & 1625.0 $\pm$ 4.2 & 44.0 $\pm$ 5.8 & 97.4 $\pm$ 1.4 & 1.9 \\
NGC 908 & 19.4 & 94.1 & 02\h23\m04\fs2 $\pm$ 0\fs05 & -21\arcdeg14\arcmin01.4\arcsec $\pm$ 0.5\arcsec & 1504.7 $\pm$ 2.6 & 54.1 $\pm$ 2.0 & 72.6 $\pm$ 1.1 & 1.8 \\
NGC 1325 & 23.7 & 115 & 03\h24\m24\fs8 $\pm$ 0\fs14 & -21\arcdeg32\arcmin45.1\arcsec $\pm$ 2.0\arcsec & 1580.8 $\pm$ 3.9 & 70.5 $\pm$ 4.2 & 54.1 $\pm$ 2.2 & 3.3 \\
NGC 1964 & 20.9 & 101 & 05\h33\m21\fs6 $\pm$ 0\fs01 & -21\arcdeg56\arcmin43.7\arcsec $\pm$ 0.5\arcsec & 1669.8 $\pm$ 1.5 & 73.6 $\pm$ 0.6 & 32.6 $\pm$ 0.5 & 1.6 \\
NGC 2280 & 24.0 & 116 & 06\h44\m49\fs0 $\pm$ 0\fs04 & -27\arcdeg38\arcmin15.2\arcsec $\pm$ 0.9\arcsec & 1873.7 $\pm$ 2.2 & 63.5 $\pm$ 1.1 & 156.3 $\pm$ 0.7 & 2.4 \\
NGC 3705 & 18.5 & 89.7 & 11\h30\m07\fs7 $\pm$ 0\fs12 & +09\arcdeg16\arcmin34.8\arcsec $\pm$ 2.8\arcsec & 1006.9 $\pm$ 4.6 & 66.1 $\pm$ 3.8 & 118.8 $\pm$ 2.1 & 3.4 \\
NGC 4517A & 26.7 & 129 & 12\h32\m28\fs1 $\pm$ 0\fs15 & +00\arcdeg23\arcmin24.0\arcsec $\pm$ 1.4\arcsec & 1488.0 $\pm$ 2.5 & 50.8 $\pm$ 4.7 & 15.8 $\pm$ 3.7 & 5.0 \\
NGC 4939 & 41.6 & 202 & 13\h04\m14\fs3 $\pm$ 0\fs03 & -10\arcdeg20\arcmin23.2\arcsec $\pm$ 0.9\arcsec & 3126.2 $\pm$ 3.3 & 56.4 $\pm$ 2.0 & 6.4 $\pm$ 0.6 & 1.9 \\
NGC 5364 & 18.1 & 87.8 & 13\h56\m11\fs4 $\pm$ 0\fs33 & +05\arcdeg00\arcmin47.5\arcsec $\pm$ 2.5\arcsec & 1249.5 $\pm$ 4.3 & 45.1 $\pm$ 6.5 & 36.6 $\pm$ 1.9 & 2.0 \\
NGC 6118 & 22.9 & 111 & 16\h21\m48\fs3 $\pm$ 0\fs06 & -02\arcdeg16\arcmin59.9\arcsec $\pm$ 1.0\arcsec & 1570.3 $\pm$ 2.7 & 67.2 $\pm$ 1.9 & 50.3 $\pm$ 1.4 & 1.4 \\
NGC 6384 & 19.7 & 95.5 & 17\h32\m24\fs4 $\pm$ 0\fs08 & +07\arcdeg03\arcmin40.8\arcsec $\pm$ 1.3\arcsec & 1682.0 $\pm$ 2.0 & 55.0 $\pm$ 2.8 & 30.7 $\pm$ 0.9 & 3.0 \\
NGC 7606 & 34.0 & 165 & 23\h19\m04\fs6 $\pm$ 0\fs02 & -08\arcdeg29\arcmin05.0\arcsec $\pm$ 0.4\arcsec & 2247.8 $\pm$ 1.6 & 66.2 $\pm$ 0.8 & 144.9 $\pm$ 0.3 & 1.4 \\
NGC 7793 & 3.44 & 16.7 & 23\h57\m50\fs6 $\pm$ 0\fs38 & -32\arcdeg35\arcmin32.6\arcsec $\pm$ 4.5\arcsec & 220.1 $\pm$ 3.6 & 39.8 $\pm$ 6.3 & 99.2 $\pm$ 6.3 & 4.7 \\
\enddata
\tablecomments{The parameters of our best-fitting axisymmetric \df\ models. From left to right, columns are: (1) galaxy name, (2-3) distance and angular scale reproduced from Table \ref{tab:tab1}, (4-5) right ascension and declination of the galaxy center, (6) systemic velocity, (7) inclination, (8) position angle, and (9) reduced-$\chi^2$ for the best fitting model.}
\end{deluxetable*}

We fit for these 6 parameters simultaneously using a
$\chi^2$-minimization routine, where the uncertainties in the pixel
intensities arise primarily from photon shot noise.  The shot noise
uncertainties are propagated through the various image reduction steps
(flattening, normalization, sky subtraction, convolution) to arrive at
a final uncertainty for the intensity at each pixel.  To account for
the uncertainty in overall normalization of each image, we also add a
small fraction of the original image intensity (typically 3-5\%) in
quadrature to the uncertainty at each pixel.

The $\chi^2$-minimization routine also returns an estimate of the
variances and covariances of our 6 model parameters.  We mask all
pixels with $\Delta F_H / F_H > 1$ or $\Delta \sigma_g / \sigma_g > 1$
to ensure that only pixels with sufficiently well-constrained
parameters are retained.  Here $\Delta$ refers to the
$\chi^2$-estimated uncertainty in a parameter.

Figure \ref{fig:lineprofs} shows an assortment of spectra and line
profile fits from our data cubes ranging from very high
signal-to-noise regions (left column) to very low signal-to-noise
regions (right column).  The line profiles shown are the best fits to
all of the data points from multiple observations combined into a
single data cube.

A number of other groups \citep[e.g.][]{2003MNRAS.342..345C,
  2015MNRAS.451.1004E} use Voronoi binning to combine pixels with low
S/N in order to bring out possible faint emission.  We have decided
not to do that.\Ignore{ because we would not then know the precise sky
  position of any faint signal that this procedure finds, which would
  complicate, and possibly throw off, our attempts to fit the rotation
  curve.}

In converting wavelengths to velocities, we first adjust our
wavelengths to the rest frame of the host galaxy by using the systemic
velocities in Table \ref{tab:tab2}.  We then use the relativistic
Doppler shift equation:
\begin{equation}
    v = c\frac{(\lambda/\lambda_0)^2-1}{(\lambda/\lambda_0)^2+1}.
\end{equation}

\subsection{Idiosyncrasies of Individual Observations}

\subsubsection{NGC 7793 Sky Subtraction}
The nearest galaxy in our sample, NGC 7793, required us to modify
slightly our procedure for subtracting the night sky emission lines
from our images.  Because it is so close, its systemic velocity is
small enough to be comparable to its internal motions; i.e.\ some of
its gas has zero line-of-sight velocity relative to Earth.
Additionally, it takes up a substantially larger fraction of the RSS
field of view than do the other galaxies discussed in this work.  This
means that night sky emission of \Ha\ and \N2 is sometimes both
spatially and spectrally coincident with NGC 7793's \Ha\ and \N2
emission across a large fraction of our images.  Because the night sky
emission was contaminated by the emission from NGC 7793, we were
unable to use the ``fit-and-subtract'' technique as described in
\S \ref{sec:skysub}.  Instead, we temporarily masked regions of
our images in which the night sky emission ring overlapped the galaxy
and fit only the uncontaminated portion of the images.  Visual
inspection of the images after this process indicates that the night
sky emission was removed effectively without over-subtracting from the
galaxy's emission.

We were unable to obtain all of our requested observations of NGC 7793
before the decommissioning of SALT's medium-resolution \FP\ etalon in
2015.  Consequently, we have acquired 4 observations of the eastern
portion of this galaxy but only 1 observation of the western portion.
We are therefore able to detect \Ha\ emission from areas of lower
signal on the eastern side of the galaxy only.  All 5 observations
overlap in the central region, which is the area of greatest interest
to our survey.

\begin{figure*}
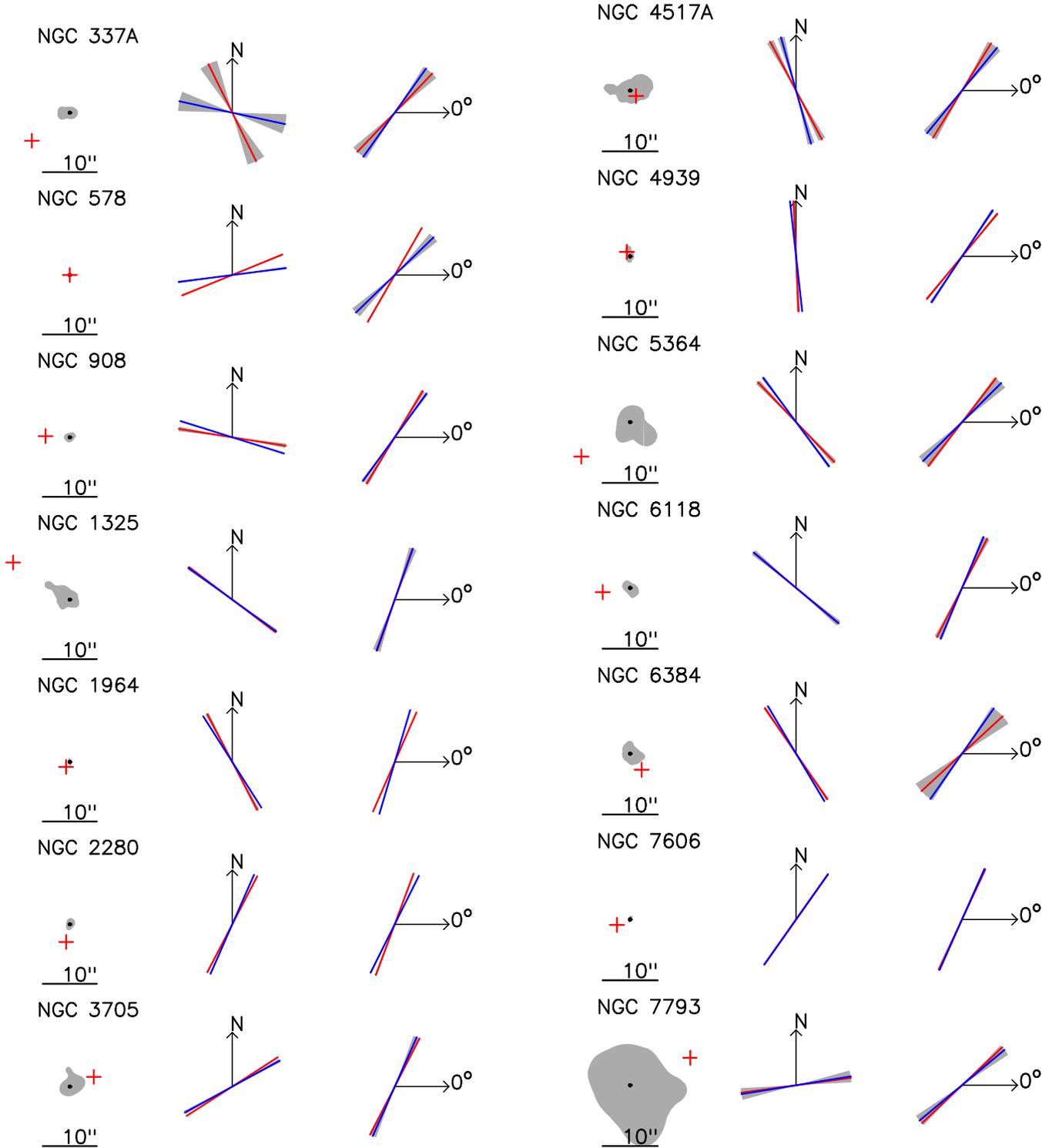

  \begin{center}
\hbox to \hsize{
    \includegraphics[width=.45\hsize]{projcomp1.ps} \hfill
    \includegraphics[width=.45\hsize]{projcomp2.ps}}
  \end{center}
  \caption{Comparison between the projection geometry fitted to the
    kinematic map (blue) and the {\it I}-band photometric image (red).
    For each galaxy, the left-hand panel compares the fitted positions
    of the centers, with the shaded area showing the region that
    encloses 68\% of the bootstrap estimates of the position of the
    best fit kinematic center, which is marked by the blue dot.  The
    red plus symbol shows the location of the adopted photometric
    center on the same scale, for which there is no uncertainty
    estimate.  Note that the center of NGC~578 was fixed at the
    photometric position when fitting the kinematic map.  The fitted
    PA is shown in the middle panel and the inclination in the
    righthand panel, and again the gray shading indicates the
    1-$\sigma$ uncertainties about the best fit value, which is less
    than the line width in some cases.
    \label{fig:projcomp}}
\end{figure*}

\subsubsection{Migratory Image Artifacts}
\label{sec:ufos}
In our 28 Dec 2011 observations of NGC 908, NGC 1325, and NGC 2280 and
our 29 Dec 2011 observation of NGC 578, we detect a series of bright
objects which move coherently across our images.  These objects have a
different point spread function from that of the real objects in our
images, and appear to be unfocused.  In a time sequence of images,
these objects move relative to the real objects of the field in a
uniform way.

The relative abundance of these objects appears to be roughly
proportional to the abundance of stars in each image, though we have
been unable to register these objects with real stars.  In the case of
our 29 Dec 2011 observation of NGC 578, one of these objects is so
bright that its diametric ghost (see \S \ref{sec:ghosts}) is
visible and moves in the opposite direction to the other objects'
coherent movement.

Based on this information, we have arrived at a possible explanation
for the appearance of these strange objects.  We believe that on these
two nights in Dec 2011, a small subset of SALT's segmented primary
mirror, perhaps only one segment, was misaligned with the rest of the
primary mirror.  This subset of the primary mirror then reflected
out-of-field light into our field.  As the secondary optics package
moved through the focal plane to track our objects of interest, the
stars reflected from outside the field then appear to move across the
images due to the misalignment of this subset of mirror segments.  New
edge sensors have been installed between SALT's primary mirror
segments in the time since these observations were taken, so these
types of image artifacts should not be present in future observations.

We have applied a simple mask over our images wherever these objects
appear.  Any pixels which fall within this mask are excluded from any
calculations in the remainder of our data reduction process.

\subsubsection{Other Image Artifacts}
SALT utilizes a small probe to track a guide star over the course of
an observation to maintain alignment with a target object.  In some of
our observations, the shadow of this guide probe overlaps our images
(e.g.\ the lower right of the image in Figure \ref{fig:ghosts}).
Similar to our treatment of the migrating objects above, we apply a
mask over pixels which are affected by this shadow.  We also apply
such a mask in the rare cases in which a satellite trail overlaps our
images.

\section{Velocity and Intensity Maps}
\label{sec:dataproduct}
The results of the foregoing reductions of the raw data cube for each
galaxy are 2D maps of median surface brightness, continuum surface
brightness (i.e.\ $C$ from equation \ref{eqn:model}), integrated
\Ha\ line surface brightness ($F_H$), integrated \N2 line surface
brightness ($F_N$), line-of-sight velocity, and estimated uncertainty
in velocity for each of our 14 galaxies.  The total number of fitted
pixels and number of independent resolution elements in each galaxy's
maps are summarized in Table \ref{tab:tab1}.

\subsection{Axisymmetric models and rotation curves}
\label{sec:models}
We have utilized the \df\footnote{\df\ is publicly available for
  download at https://www. physics.queensu.ca/Astro/people/Kristine\underline{~}Spekkens/diskfit/}
software package \citep{2007ApJ...664..204S,2010MNRAS.404.1733S} to
fit axisymmetric rotation models to our \Ha\ velocity fields.  Unlike
tilted-ring codes, e.g.\ \texttt{rotcur} \citep{rotcur}, \df\ assumes
a single projection geometry for the entire galactic disk and derives
uncertainties on all the fitted parameters from a bootstrap procedure.

In addition to fitting for five global parameters, which mostly
describe the projection geometry, it fits for a circular rotation
speed in each of an arbitrary number of user-specified radius bins
(i.e.\ the rotation curve).  The five global parameters are: the
position of the galaxy center ($x_c, y_c$), the systemic recession
velocity of the galaxy ($V_{\rm sys}$), the disk inclination ($i$), and
the position angle of the disk relative to the North-South axis
($\phi_{\rm PA}$).  For $N$ user-specified radius bins, \df\ fits for the
$N+5$ parameters using a $\chi^2$-minimization algorithm.

Where we have sufficiently dense velocity measurements, we typically
space the $N$ radial bins along the major axis by 5\arcsec, which well
exceeds the seeing in all cases, so that each velocity measurement is
independent.

The velocity uncertainties used in calculating the $\chi^2$ values
arise from two sources: the uncertainty in fitting a Voigt profile to
each pixel's spectrum (\S \ref{sec:linefits}) and the intrinsic
turbulence within a galaxy.  This intrinsic turbulence, $\Delta_{\rm
  ISM}$, is in the range 7-12~km~s$^{-1}$ both in the Milky Way
\citep{1979AJ.....84.1181G} and in external galaxies
\citep{1993PhDT.......185K}.  When most emission in a pixel arises
from a single \ion{H}{2} region, the measured velocity may differ from
the mean orbital speed by some random amount drawn from this turbulent
spread.  We therefore add $\Delta_{\rm ISM} = 12~$km~s$^{-1}$ in
quadrature to the estimated velocity uncertainty in each pixel when
fitting these models to each of our galaxies.

We calculate uncertainties for each of these fitted parameters using
the bootstrap method described in \citet{2010MNRAS.404.1733S}.  Due to
the fact that these velocity maps can contain structure not accounted
for in our models, residual velocities may be correlated over much
larger regions than a single resolution element.  To account for this,
the bootstrap method preserves regions of correlated residual line of
sight velocity when resampling the data to estimate the uncertainty
values.

Table \ref{tab:tab2} lists the projection parameters and
reduced-$\chi^2$ values for our best-fitting axisymmetric models to
our 14 \Ha\ velocity maps.  The uncertainty values in Table
\ref{tab:tab2} and in the rotation curves of Figures
\ref{fig:N337A}-\ref{fig:N7793} are the estimated 1-$\sigma$
uncertainties from 1000 bootstrap iterations.  In some cases
(e.g.\ NGC 7793), the inclination of the galaxy is poorly constrained
in our axisymmetric models.  This leads to a large uncertainty in the
overall normalization of the rotation curve even when the shape of the
rotation curve is well-constrained.  This is the reason that
uncertainties in the velocities are often substantially larger than
the point-to-point scatter in the individual values.

\subsection{Non-axisymmetric models}
\df\ is also capable of fitting more complicated models that include
kinematic features such as bars, warped disks, and radial flows.  We
have attempted to fit our velocity maps with such models, but in no
case have we obtained an improved fit that appeared convincing.  Often
a fitted ``bar'' was clearly misaligned with, and of different
length from that visible in the galaxy image, and the bootstrap
uncertainties yielded large errors on the fitted bar parameters.  The
\df\ algorithm has been demonstrated to work well
\citep{2007ApJ...664..204S, 2010MNRAS.404.1733S} when there are
well-determined velocities covering the region of the bar.  But the
\df\ algorithm is unable to find a convincing fit when the velocity
map lacks information at crucial azimuths of the expected bar flow, as
appears to be the case for all the barred galaxies in our sample.
This remains true even when the initial guesses at parameter values
are chosen carefully.  We therefore here present only axisymmetric
fits to our data in which bars and other asymmetries are azimuthally
averaged.  We will discuss more complex kinematic models for these
galaxies in future papers in this series.

\subsection{Comparison with photometry}
\citet{RINGSPhot}\ have applied the \df\ package to multi-band
photometric images of these galaxies, fitting both a disk and, where
appropriate, a bulge and/or a bar.  These fits yield the disk major
axis position angle and an axis ratio that is interpreted as a measure
of the inclination of a thin, round disk.  In order to estimate color
gradients, they fixed the photometric center in each image to the same
sky position, and therefore did not obtain uncertainty estimates for
the position of the center.  Figure~\ref{fig:projcomp} presents a
graphical comparison between the values derived separately from our
kinematic maps and from the {\it I}-band image of each galaxy.  In
most cases, the measurements agree within the uncertainties.  However,
there are some significant differences.  In particular, discrepancies
in the fitted positions of the centers seem large compared with the
uncertainties.  In some cases, notably NGC 1325, NGC 3705, NGC 5364,
NGC 6384, and NGC 7606, we have no kinematic measurements in the inner
15\arcsec\ - 25\arcsec, which complicates fitting for the center.  In
all these cases, both the kinematic and photometric centers are well
within the region where we have no kinematic data, while the radial
extent of our maps is 10 - 20 times larger; forcing the kinematic
center to coincide with the photometric center has little effect on
the fitted inclination, position angle, and outer rotation curve.  We
discuss other cases in the following subsections about each galaxy

\begin{figure*}[t]
  \begin{center}
    \includegraphics[width=0.9\hsize]{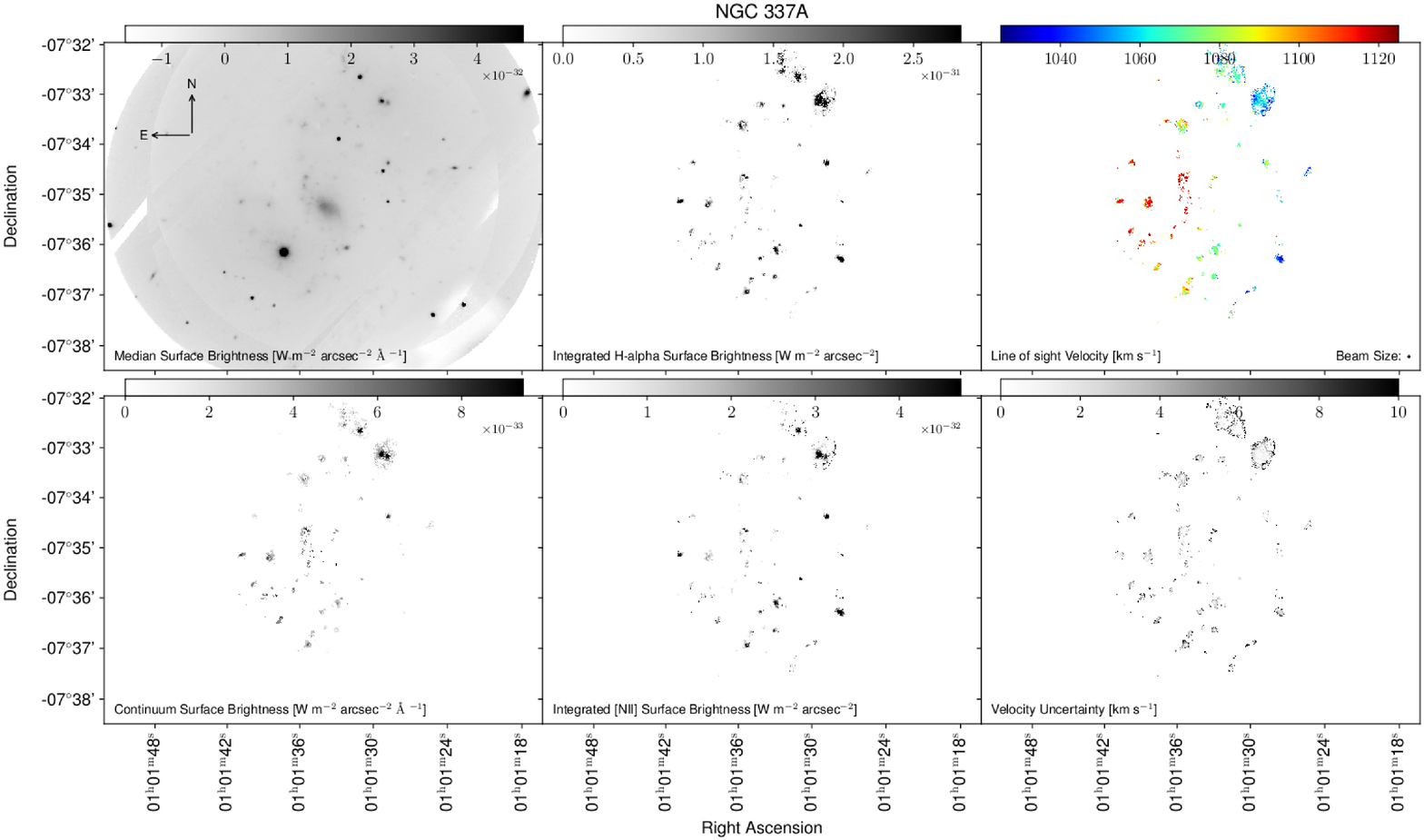}
    \includegraphics[width=\hsize]{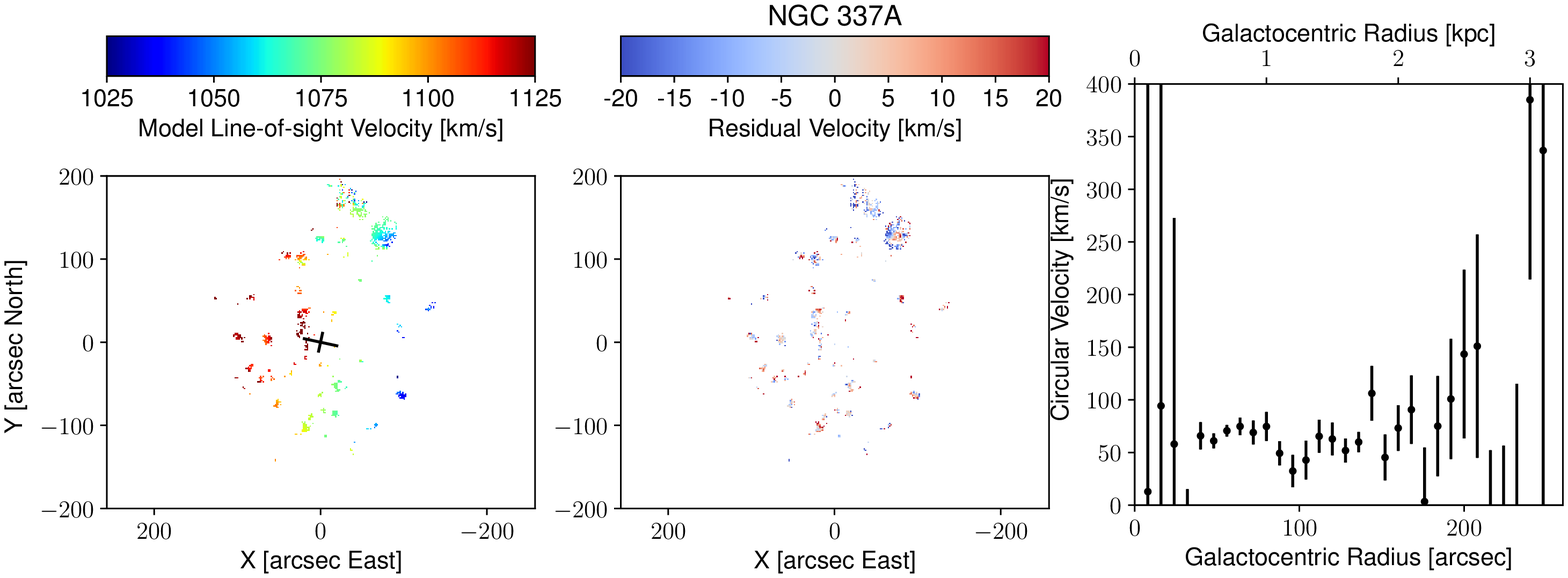}
  \end{center}
  \caption{Results for NGC 337A.  Top left: the median flux for each
    pixel in our combined data cube.  Middle left: the fitted
    continuum flux.  Top center: the fitted integrated \Ha\ line flux.
    Middle center: the fitted integrated \N2 line flux.  Top right:
    the fitted line-of-sight velocity.  Middle right: the estimated
    uncertainty in the fitted line-of-sight velocity.  At a distance
    of 2.57 Mpc, the physical scale is $12.5 \textrm{ pc}/\arcsec$.
    Bottom left: Our best-fitting axisymmetric \df\ model of NGC
    337A's line-of-sight \Ha\ velocity field. The center, orientation
    of the major axis, and axis-ratio of our best-fitting \df\ model
    are marked with a large black cross.  Bottom center: A map of the
    data-minus-model residual velocities for the best-fitting model in
    the left panel.  Bottom right: A rotation curve extracted from the
    best-fitting axisymmetric model with 1-$\sigma$ uncertainties
    derived from our bootstrapping procedure.  The radii were chosen
    to be at least 5\arcsec\ apart, which is approximately 2 seeing
    elements.
    \label{fig:N337A}}
\end{figure*}

\section{Results for Individual Galaxies}
\subsection{NGC 337A}
NGC 337A has one of the most sparsely sampled velocity maps in the
RINGS medium-resolution \Ha\ kinematic data, as seen in Figure
\ref{fig:N337A}.  It is also one of the two galaxies in this work
(along with NGC 4517A) that are classified as Irregular.  Despite
this, our model is able to sample the rotation curve over a wide range
of radii (Figure \ref{fig:N337A}) extending out to $\sim 2.5~$kpc.
Near the center and at $R\ga 175$\arcsec, the velocity data are too
sparse to yield a meaningful estimate of the circular speed.

\begin{figure*}
  \begin{center}
    \includegraphics[width=0.9\hsize]{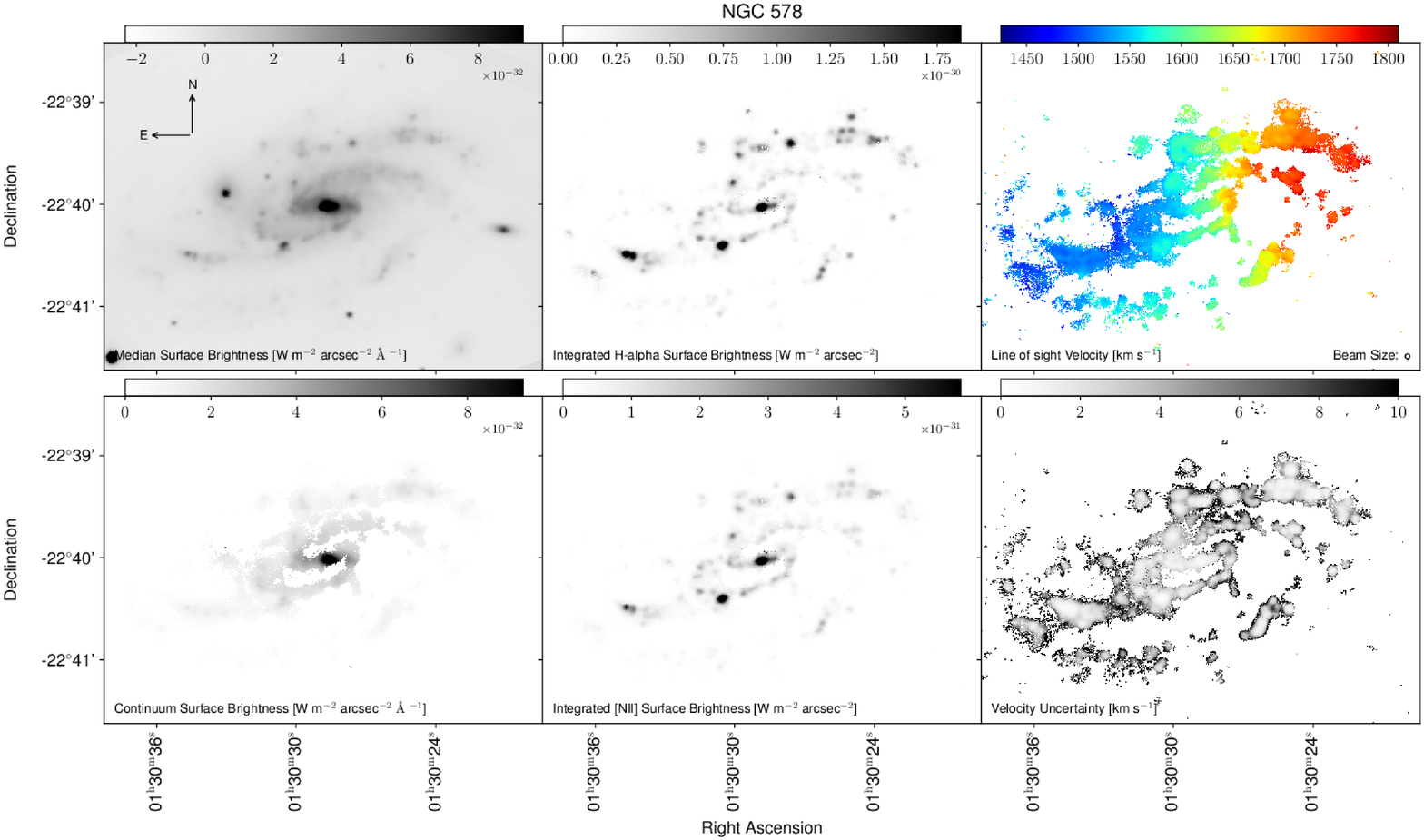}
    \includegraphics[width=\hsize]{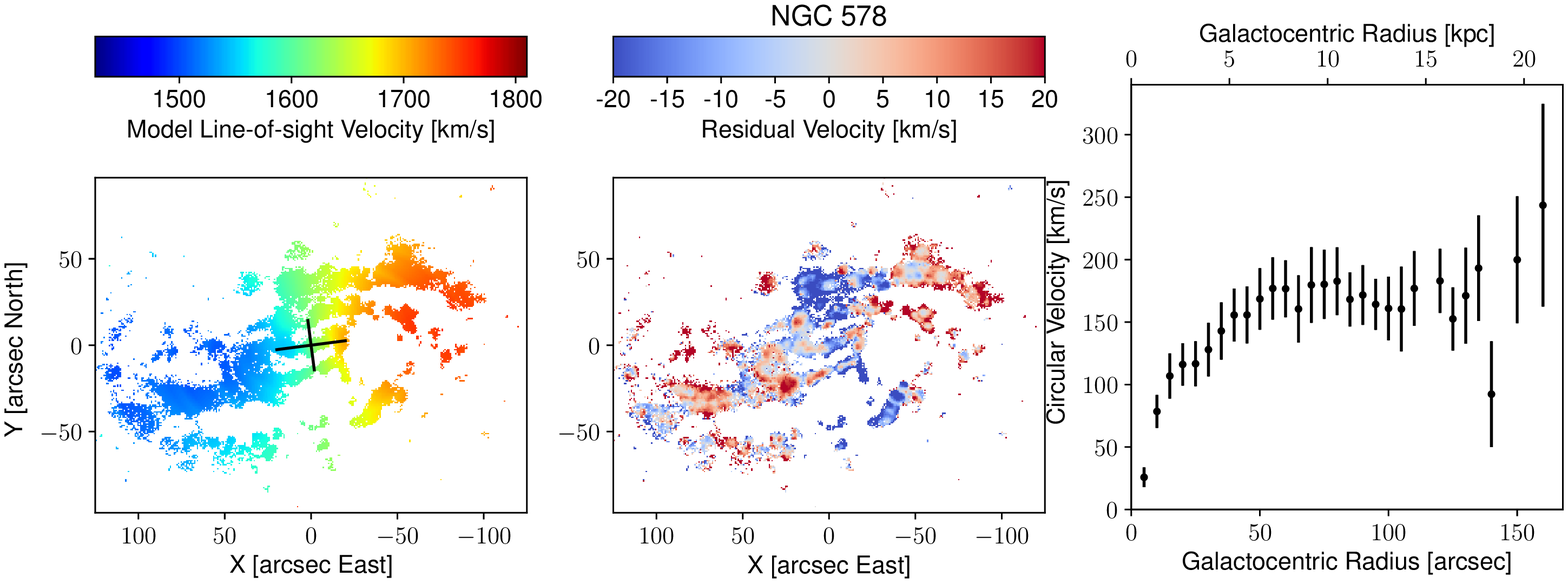}
  \end{center}
  \caption{Same as Figure \ref{fig:N337A}, but for NGC 578.  At a
    distance of 27.1 Mpc, the physical scale is $131 \textrm{
      pc}/\arcsec$.  The large uncertainties on the points are due
    almost entirely to the galaxy's inclination being poorly
    constrained.
    \label{fig:N578}}
\end{figure*}


Our best-fitting kinematic projection parameters for this galaxy differ
substantially from those derived from the {\it I}-band image by
\citet{RINGSPhot}, as indicated in Figure~\ref{fig:projcomp}, which is
not too surprising given the sparseness of the kinematic map.  In
particular, the axis about which the galaxy is rotating appears to be
strongly misaligned from the symmetry axis of the {\it I}-band light
distribution.  Since the kinematic data are clearly blueshifted on the
West side of the galaxy and redshifted on the East, the misalignment
is more probably due to difficulties in fitting the image; the light
of NGC~337A is dominated by a bulge while the disk is very faint so
that the apparent projection geometry of the galaxy is dominated by
that of the bulge.


\subsection{NGC 578}
Even though NGC 578 exhibits one of the strongest visible bars among
this sample of galaxies, we were disappointed to find that the
velocity map (Figure~\ref{fig:N578}) lacks sufficient data in the bar
region to be able to separate a non-circular flow from the
axisymmetric part.  Note the absence of velocity information
immediately to the N and S of the bar.  We therefore derive an
estimate of the rotation curve from an axisymmetric fit only.  Also,
for this galaxy only, we fix the center of rotation to the sky
position of the photometric center.  The coherent velocity features in
the residual map clearly contain more information that we will examine
more closely in a future paper in this series.

The slow, and almost continuous rise of the fitted circular speed
affects our ability to determine the inclination of the disk plane to
line of sight, which is generally more tightly constrained when the
rotation curve has a clear peak.  This galaxy therefore has one of the
larger inclination uncertainties in the sample, which leads to the
large uncertainties in the deprojection of the orbital speeds and to
the fact that the point-to-point differences in the best fit values
are substantially smaller than the uncertainties.

As shown in Figure~\ref{fig:projcomp}, the best-fitting inclination
and position angle for our kinematic models of this galaxy disagree
significantly with the values derived from the photometric model of
\citet{RINGSPhot}, in which the bar was fitted separately.  There are
at least two reasons for this discrepancy: the prominent bar feature
probably does affect the estimated projection geometry derived from an
axisymmetric fit to the kinematic map and the galaxy image also
manifests a strong asymmetry in the outer parts, with an unmatched
spiral near the Northern minor axis, that complicates the fit to
photometric image.

\begin{figure*}
    \begin{center}
    \includegraphics[width=0.9\hsize]{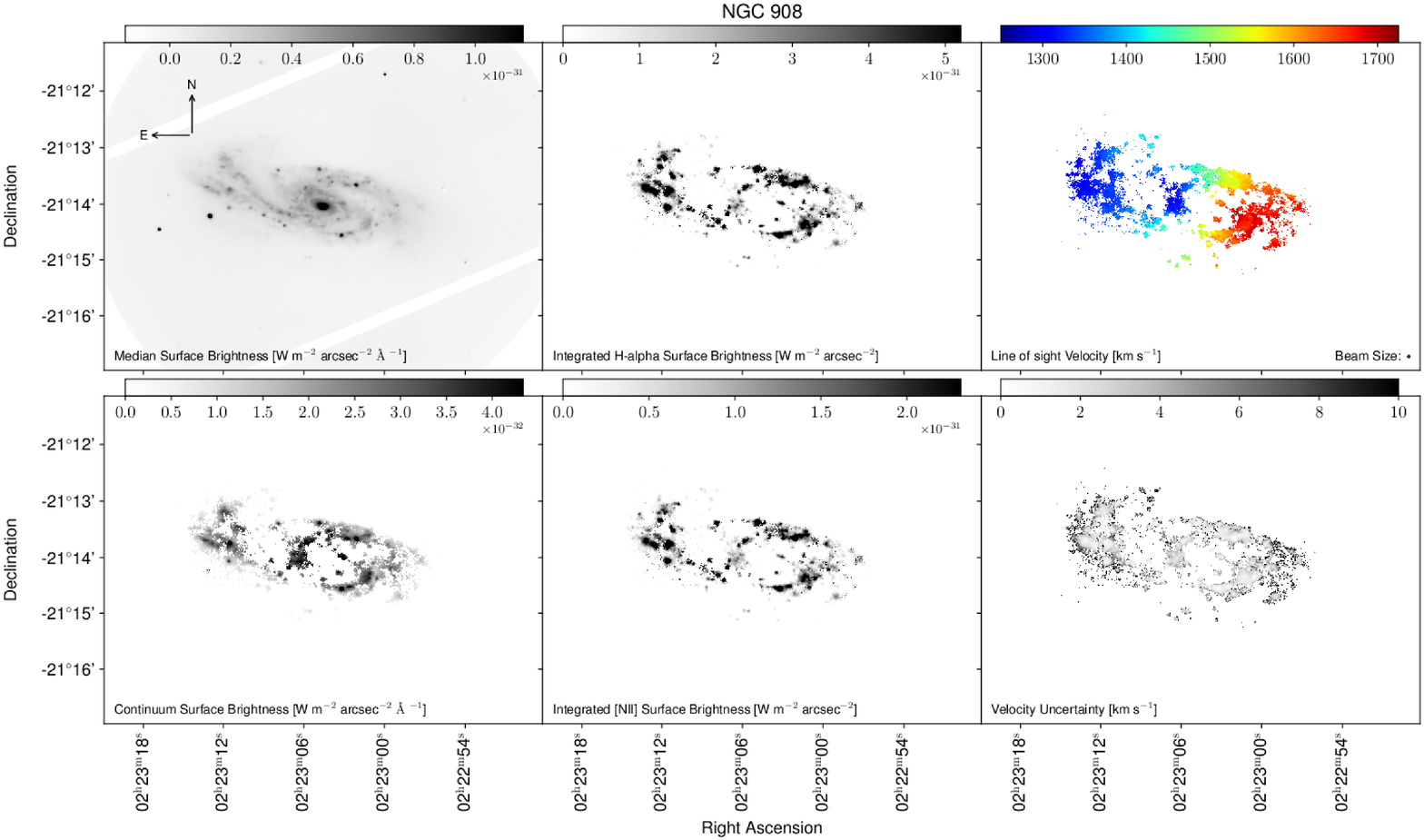}
    \includegraphics[width=\hsize]{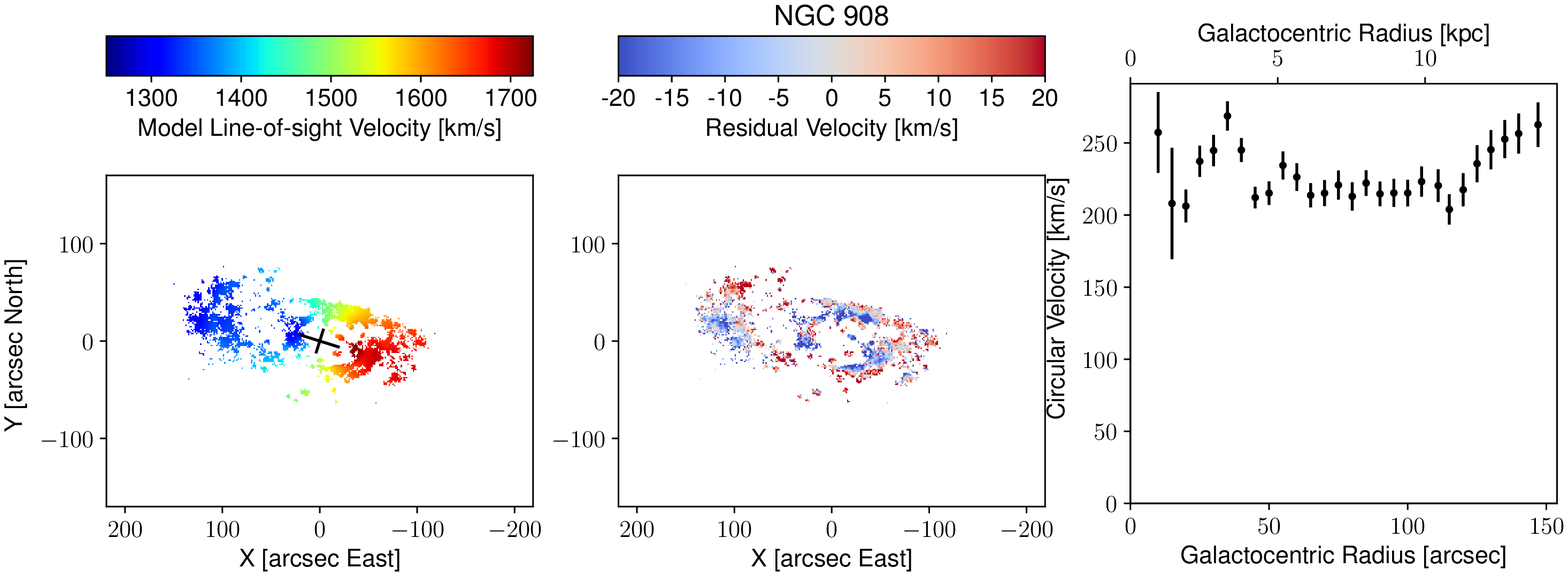}
    \end{center}
    \caption{Same as Figure \ref{fig:N337A}, but for NGC 908.  At a
      distance of 19.4 Mpc, the physical scale is $94.1 \textrm{
        pc}/\arcsec$.
    \label{fig:N908}}
\end{figure*}

In Figure~\ref{fig:rc_comp}, we compare our derived rotation curve
with that reported by \citet{1996ApJS..107...97M} via \Ha\ longslit
spectroscopy (red points).  There is generally somewhat smaller
scatter in our points, and those authors adopt a higher inclination of
58\arcdeg, compared with our 44\arcdeg, causing them to derive
circular speeds that are systematically lower by about 20\%.


\subsection{NGC 908}
NGC 908 has a single large spiral arm towards the north-east side of
the galaxy (see the top left panel of Figure \ref{fig:N908}) which is
unmatched by a corresponding spiral arm on the opposite side.  We have
fitted an axisymmetric model, which therefore leads to a corresponding
region of large correlated residual velocity.  This feature is
probably responsible for the sudden increase in the derived rotation
curve beyond 120\arcsec, which could also be indicative of a warped
disk at large radii.

Again, Figure~\ref{fig:projcomp} indicates that our best-fitting
values for the center, position angle, and inclination of this galaxy
differ somewhat from those fitted to the {\it I}-band image
\citep{RINGSPhot}, though this is not entirely surprising given the
asymmetry of this galaxy.

As shown in Figure \ref{fig:rc_comp}, the shape of our derived
rotation curve for NGC 908 agrees fairly well with the previous
long-slit measurements by \citet{1996ApJS..107...97M}, although we do
not reproduce the slow inner rise that they report.  Again they
adopted a higher inclination of 66\arcdeg, compared with our
54\arcdeg, causing their circular speeds to be lower than ours by
about 12\%.

\begin{figure*}
  \begin{center}
    \includegraphics[width=0.9\hsize]{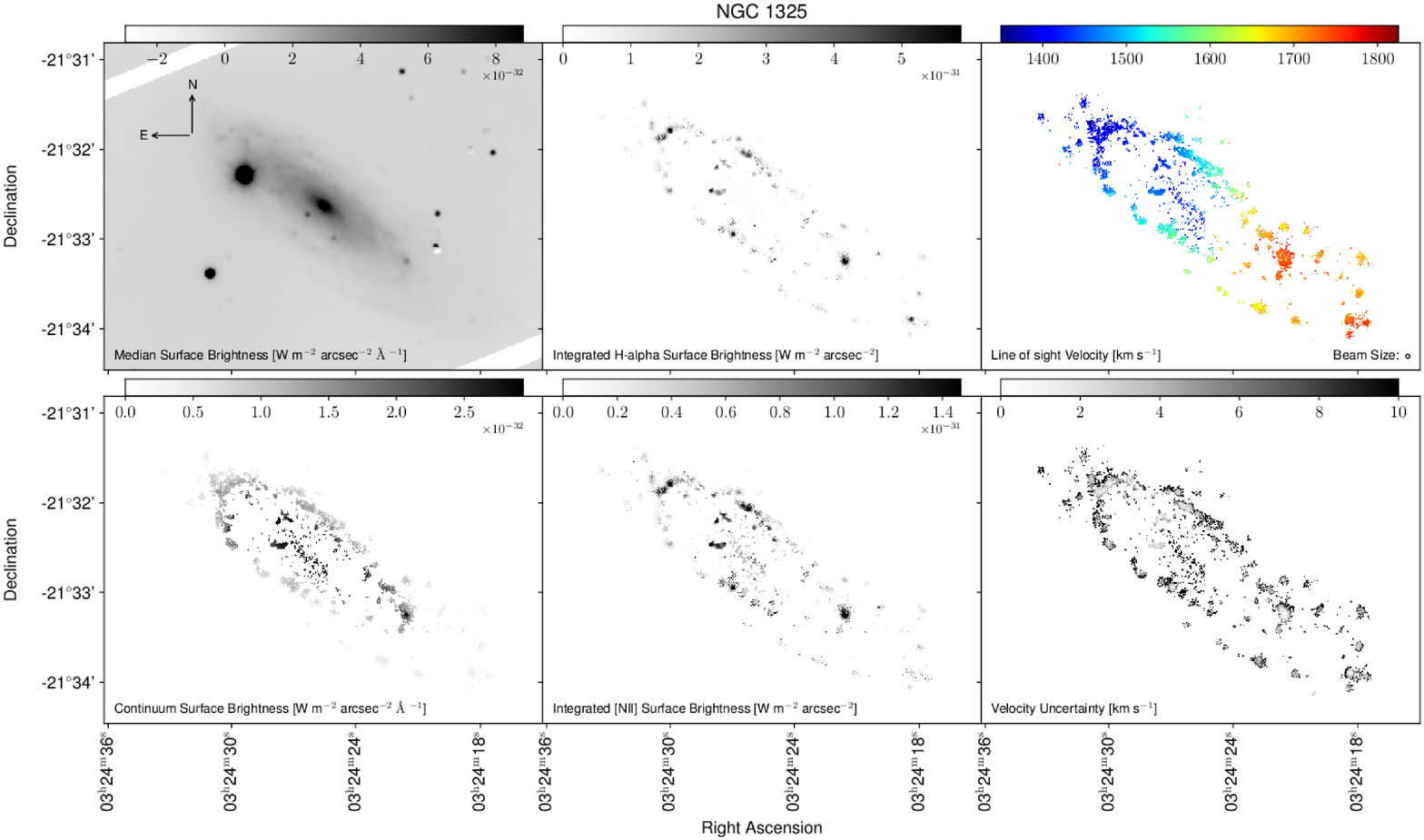}
    \includegraphics[width=\hsize]{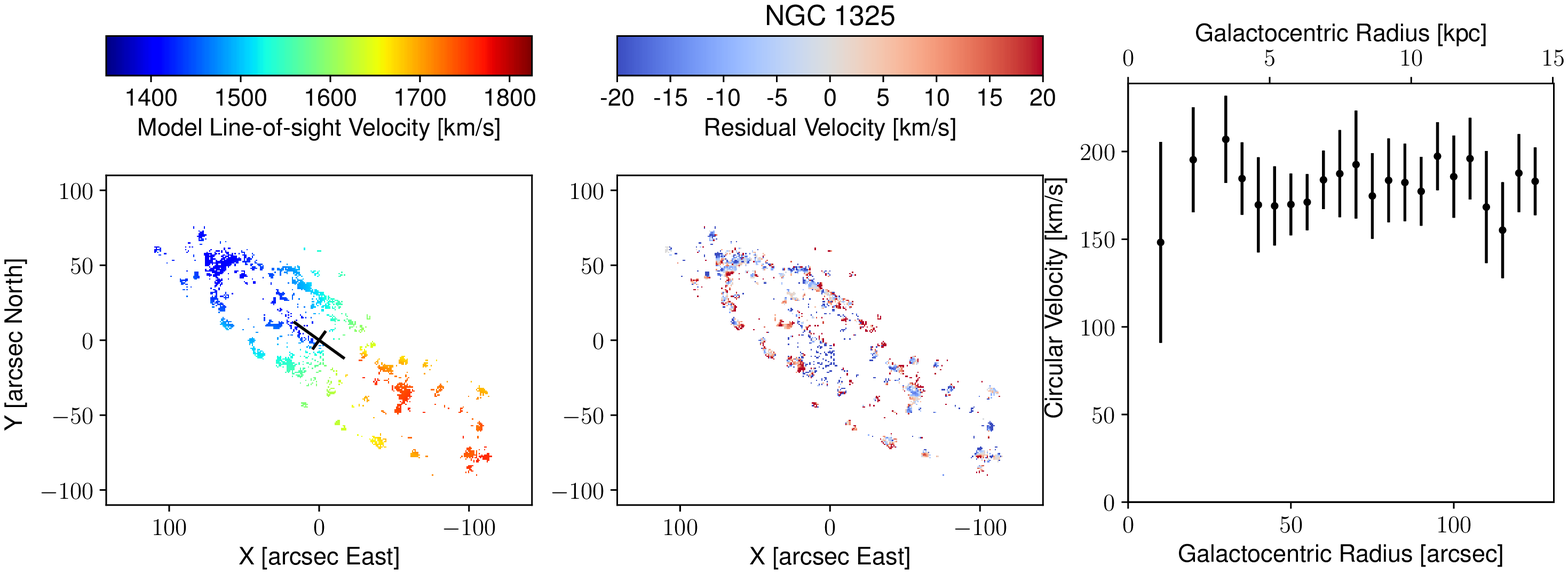}
  \end{center}
  \caption{Same as Figure \ref{fig:N337A}, but for NGC 1325.  At a
    distance of 23.7 Mpc, the physical scale is $115 \textrm{
      pc}/\arcsec$.
    \label{fig:N1325}}
\end{figure*}

\vfill\eject

\subsection{NGC 1325}
Our data on NGC 1325 (Figure \ref{fig:N1325}) indicate that this
galaxy has a regular projected flow pattern.  We derive a rotation
curve that is approximately flat over a wide range of radii.  Notably,
we detect very little \Ha\ emission in the innermost
$\sim25$\arcsec\ of the map, where our velocity estimates are
correspondingly sparse and uncertain.  Our best-fitting projection
angles for this galaxy agree extremely well with those from the
photometric models of \citet{RINGSPhot}, as shown in
Figure~\ref{fig:projcomp}, but the position of the center differs by
over 10\arcsec, probably because of the dearth of kinematic data in
the inner parts.

\citet{1982ApJ...261..439R} adopted an inclination of 70\arcdeg\ for
this galaxy, which is identical within the uncertainty with our
best fit value, and our extracted rotation curve agrees
reasonably well (Figure \ref{fig:rc_comp}) with their measurements at
$R > 50\arcsec$.  We do not, however, reproduce the slow rise
interior to this radius that they report; this discrepancy could
indicate that their slit did not pass through the center.

\begin{figure*}
  \begin{center}
    \includegraphics[width=0.9\hsize]{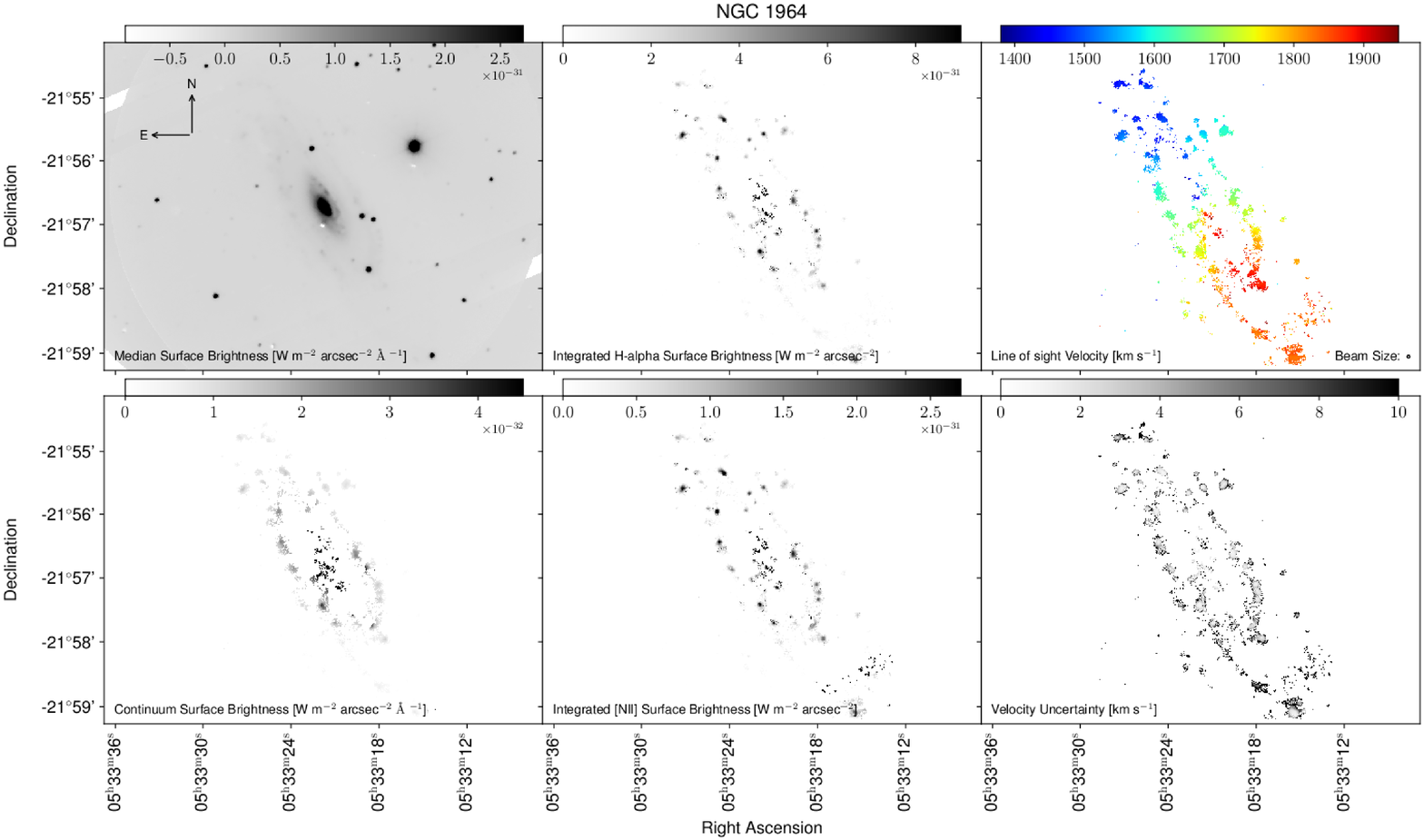}
    \includegraphics[width=\hsize]{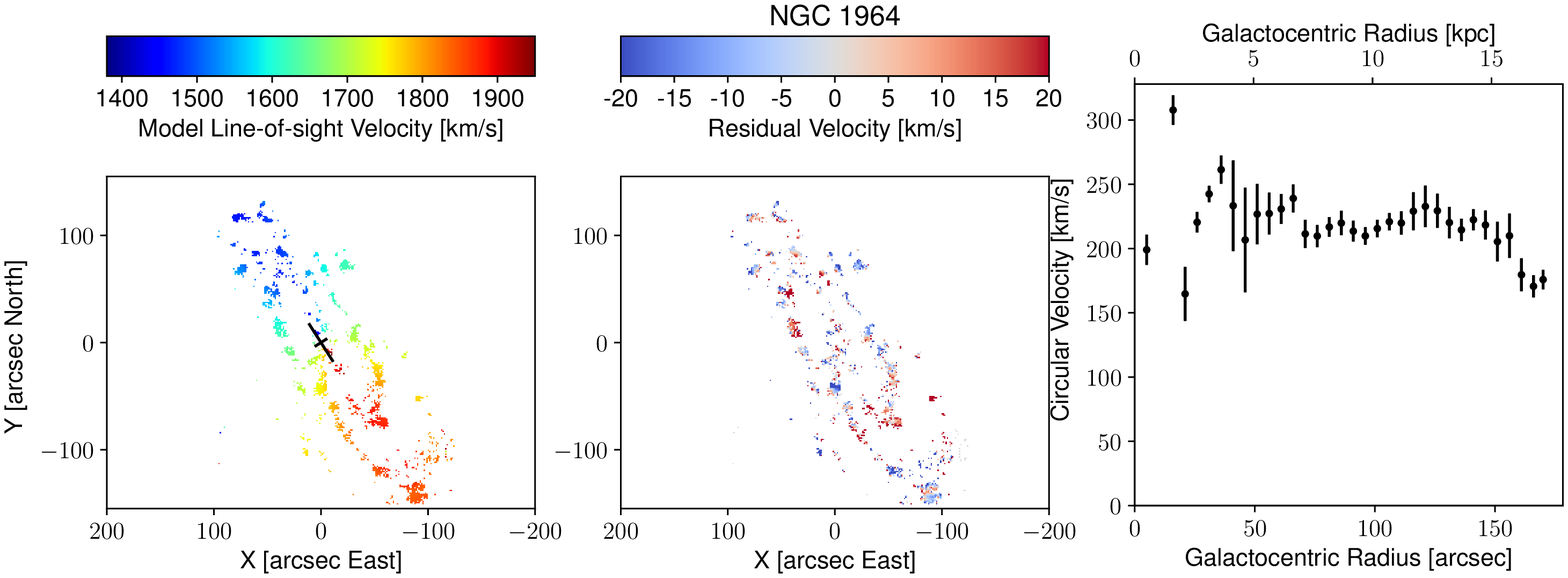}
  \end{center}
  \caption{Same as Figure \ref{fig:N337A}, but for NGC 1964.  At a
    distance of 20.9 Mpc, the physical scale is $101 \textrm{
      pc}/\arcsec$.
    \label{fig:N1964}}
\end{figure*}

\vfill\eject

\subsection{NGC 1964}
We find, Figure \ref{fig:N1964}, an almost regular flow pattern
for NGC 1964.  Our fitted center position and projection angles agree,
within the estimated uncertainties (see Figure~\ref{fig:projcomp}),
with those derived from the {\it I}-band image by \citet{RINGSPhot}.

As shown in Figure \ref{fig:rc_comp}, our derived rotation curve is
similar to that measured previously by \citet{1996ApJS..107...97M},
who adopted an inclination of 68\arcdeg, compared to our 74\arcdeg.

\begin{figure*}
  \begin{center}
    \includegraphics[width=0.9\hsize]{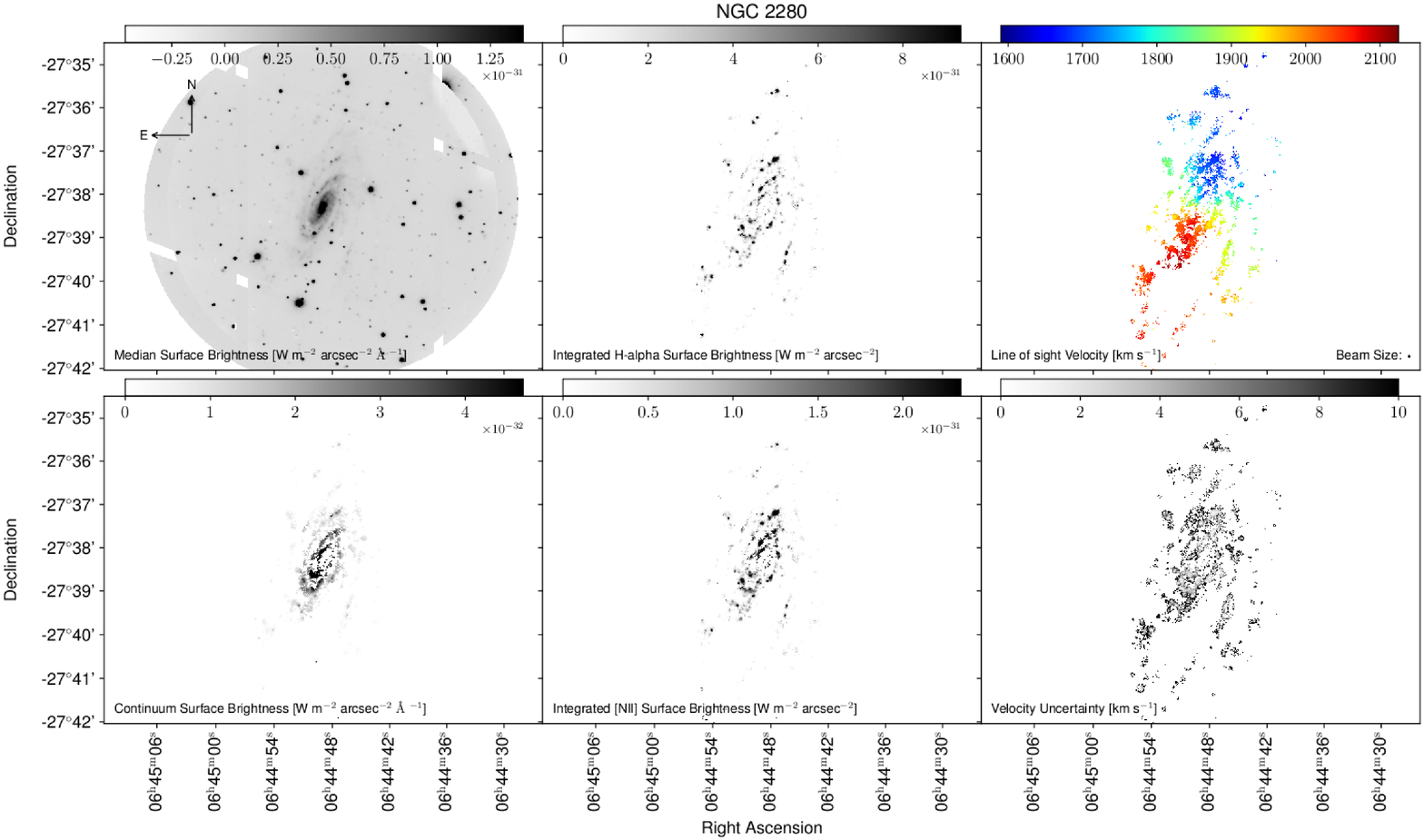}
    \includegraphics[width=\hsize]{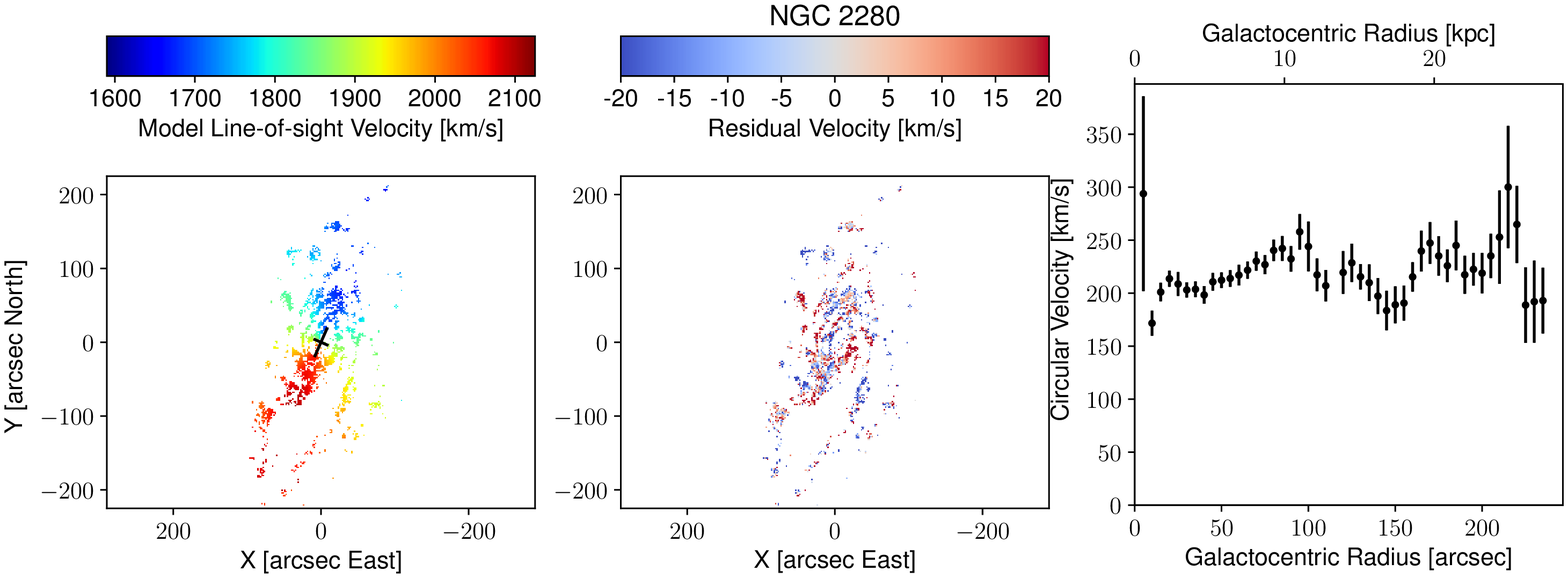}
  \end{center}
  \caption{Same as Figure \ref{fig:N337A}, but for NGC 2280.  At a distance of 24.0 Mpc, the physical scale is $116 \textrm{ pc}/\arcsec$.
    \label{fig:N2280}}
\end{figure*}

\vfill\eject\phantom{blah}\vfill\eject

\subsection{NGC 2280}
Our previous paper \citep{RINGS1} presented a kinematic map for NGC
2280 that was derived from the same \FP\ data cube.  The most
significant difference between the maps and models presented in that
work and those presented here is increased spatial resolution due
to a change in our pixel binning procedure.  As mentioned in \S\S
\ref{sec:flattening} and \ref{sec:ghosts}, we have made minor
refinements to our flatfielding and ghost subtraction routines which
have improved the data reduction process, and here we also include a
fit to the \N2 6583 line in addition to the \Ha\ line, which results in
a slightly increased image depth.

Our derived velocity map for NGC 2280, presented in
Figure~\ref{fig:N2280}, again reveals a regular flow pattern that is
typical of a rotating disk galaxy seen in projection.  Unlike many of
the other galaxies in our sample, we have been able to extract
reliable velocities at both very small and very large radii, producing
one of the most complete rotation curves in this sample.  Aside from
the inermost point, which is very uncertain, the measured orbital
speed agrees well with that in our previous paper, where we also
demonstrated general agreeement with the \ion{H}{1} rotation curve.

The position of the center, inclination, and position angle of this
galaxy are very well constrained in our models, with uncertainties
$\la 1$\arcdeg\ for both angle parameters.  These values are
consistent with our previous work on this galaxy in \citet{RINGS1},
but the estimated inclination, 63.5\arcdeg\ is in tension (see
Figure~\ref{fig:projcomp}) with the 69.6\arcdeg\ derived from the {\it
  I}-band image by \citet{RINGSPhot}, who also estimated the
uncertainty on each angle to be $\sim 1\arcdeg$.

\begin{figure*}
  \begin{center}
    \includegraphics[width=0.9\hsize]{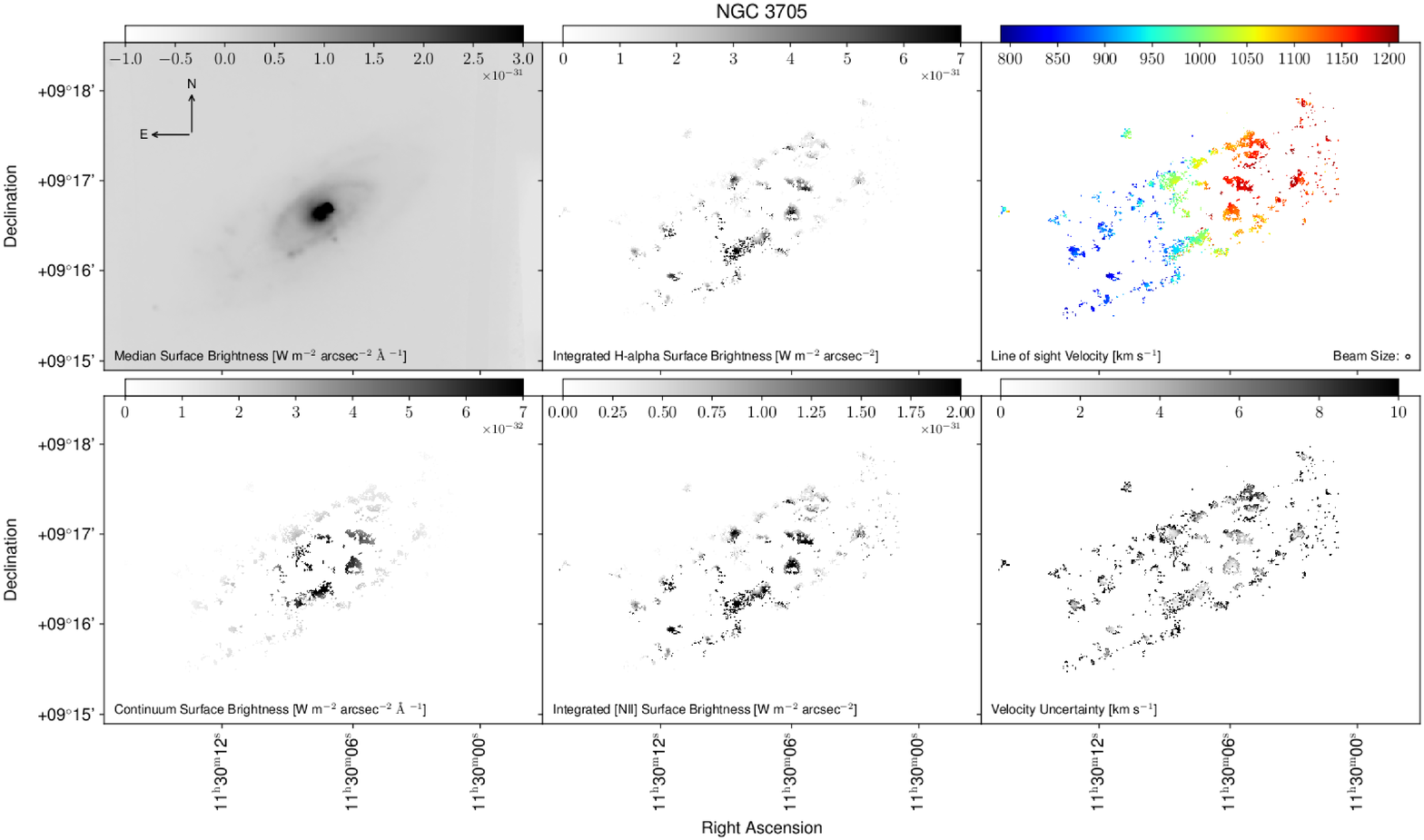}
    \includegraphics[width=\hsize]{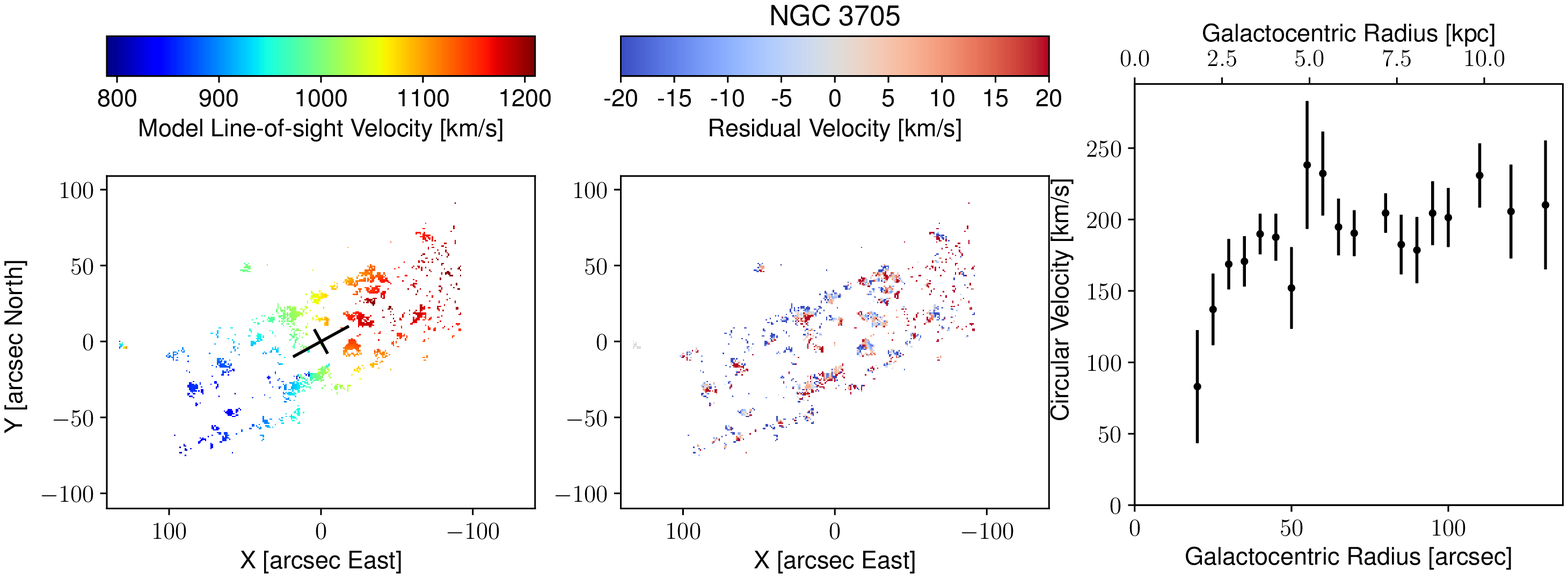}
  \end{center}
  \caption{Same as Figure \ref{fig:N337A}, but for NGC 3705.  At a
    distance of 18.5 Mpc, the physical scale is $89.7 \textrm{
      pc}/\arcsec$.
    \label{fig:N3705}}
\end{figure*}

Our rotation curve NGC 2280 extends to much larger radii than those
published previously, as shown in Figure \ref{fig:rc_comp}.  We derive
systematically slightly larger velocities than did
\citet{1996ApJS..107...97M} (red points), who adopted $i=61\arcdeg$.
Our estimated velocities are almost double the values reported by
\citet{1995A&AS..110..279S} (green points), who did not give an
inclination for this galaxy and may have reported projected, i.e.\
line-of-sight, velocities.


\subsection{NGC 3705}
We have derived the maps shown in Figure~\ref{fig:N3705} from our data
on NGC 3705.  We detect no \Ha\ emission in the central $\sim
20\arcsec$ of the galaxy, and the innermost fitted velocities have
large uncertainties.  For $R \ga 80\arcsec$, the rotation curve
appears to be approximately flat over a broad range of radii.

Our values for NGC 3705's center and projection angles are consistent
with (see Figure~\ref{fig:projcomp}) the values from the {\it I}-band image
fitted by \citet{RINGSPhot}, but our lack of velocity measurements at
small radii made it difficult to pinpoint the center.

\begin{figure*}
  \begin{center}
    \includegraphics[width=0.9\hsize]{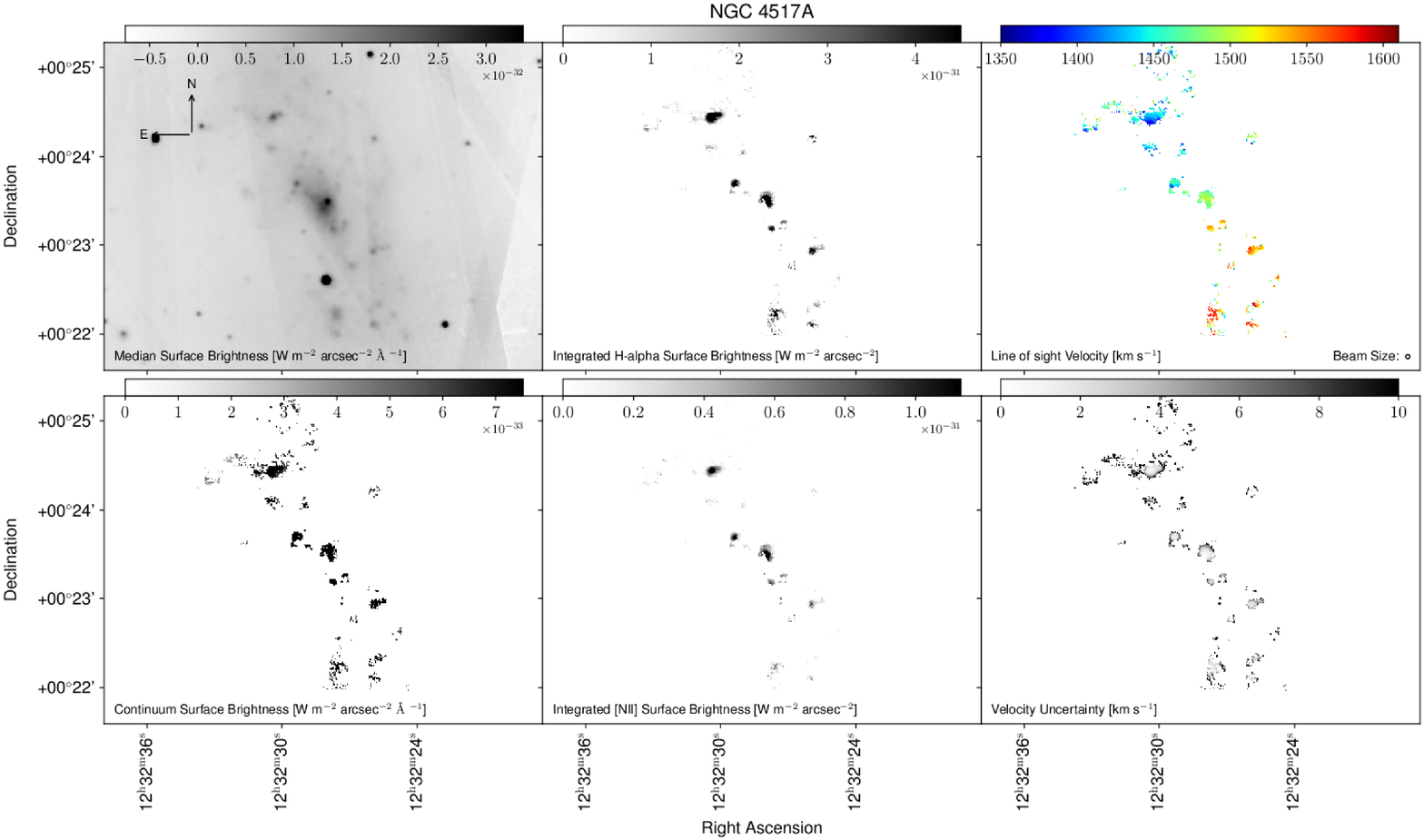}
    \includegraphics[width=\hsize]{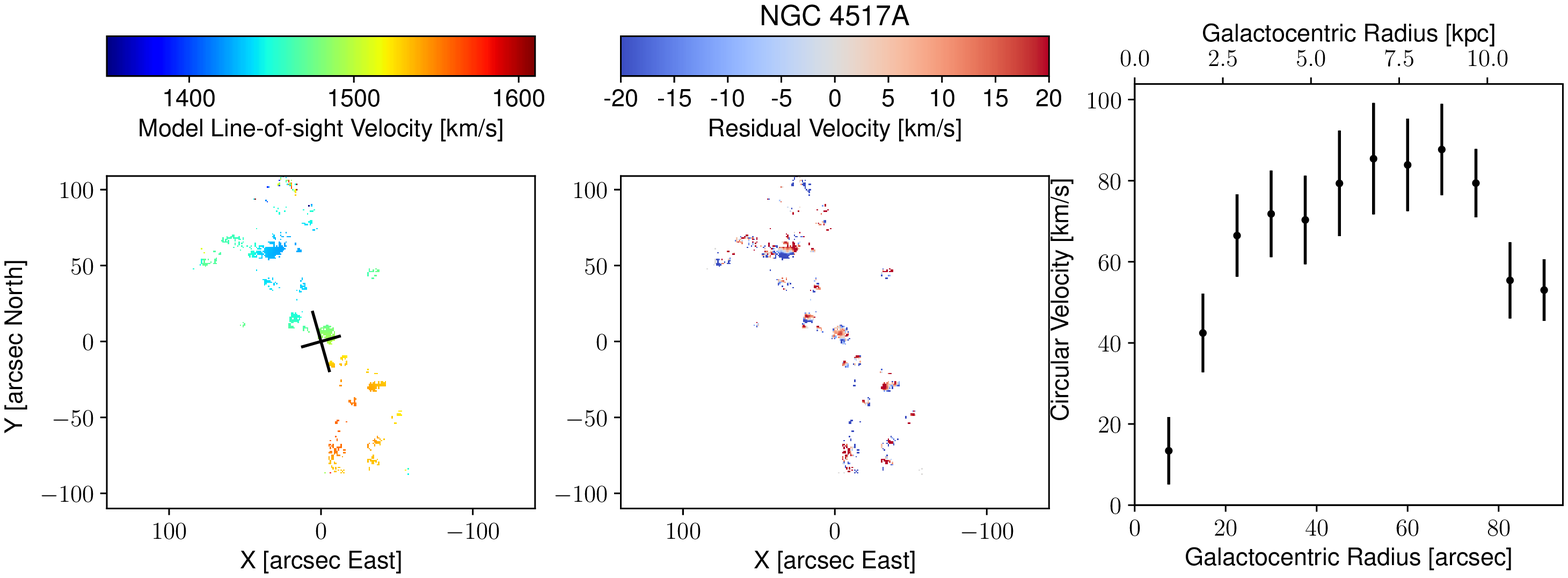}
  \end{center}
  \caption{Same as Figure \ref{fig:N337A}, but for NGC 4517A.  At a distance of 26.7 Mpc, the physical scale is $129 \textrm{ pc}/\arcsec$.
    \label{fig:N4517A}}
\end{figure*}

\vfill\eject\phantom{blah}\vfill\eject

\subsection{NGC 4517A}
Our velocity map for NGC 4517A, Figure~\ref{fig:N4517A}, like that for
NGC 337A, is very sparsely sampled, and both galaxies are
morphologically classified as Irregular.  Our rotation curve extracted
from an axisymmetric model of this galaxy is sparsely sampled and has
large uncertainties.  These uncertainties also reflect the uncertainty
in the inclination.

The projection parameters of our best-fitting model have some of the
largest uncertainties in Table \ref{tab:tab2}, but are consistent,
within the uncertainties (see Figure~\ref{fig:projcomp}), with the
values derived from the {\it I}-band image by \citet{RINGSPhot}, and our
fitted center agrees well with the photometric estimate.

Our estimates of the circular speed in NGC 4517A generally agree with
the values measured by \citet{2011MNRAS.413.1875N}, though both their
PPAK data and ours are quite sparse (Figure \ref{fig:rc_comp}).  Their
slightly higher orbital speeds are a consequence of a difference in
adopted inclination of $i = 90\arcdeg - 33\arcdeg = 57\arcdeg$
compared with our 51\arcdeg.

\begin{figure*}
  \begin{center}
    \includegraphics[width=0.9\hsize]{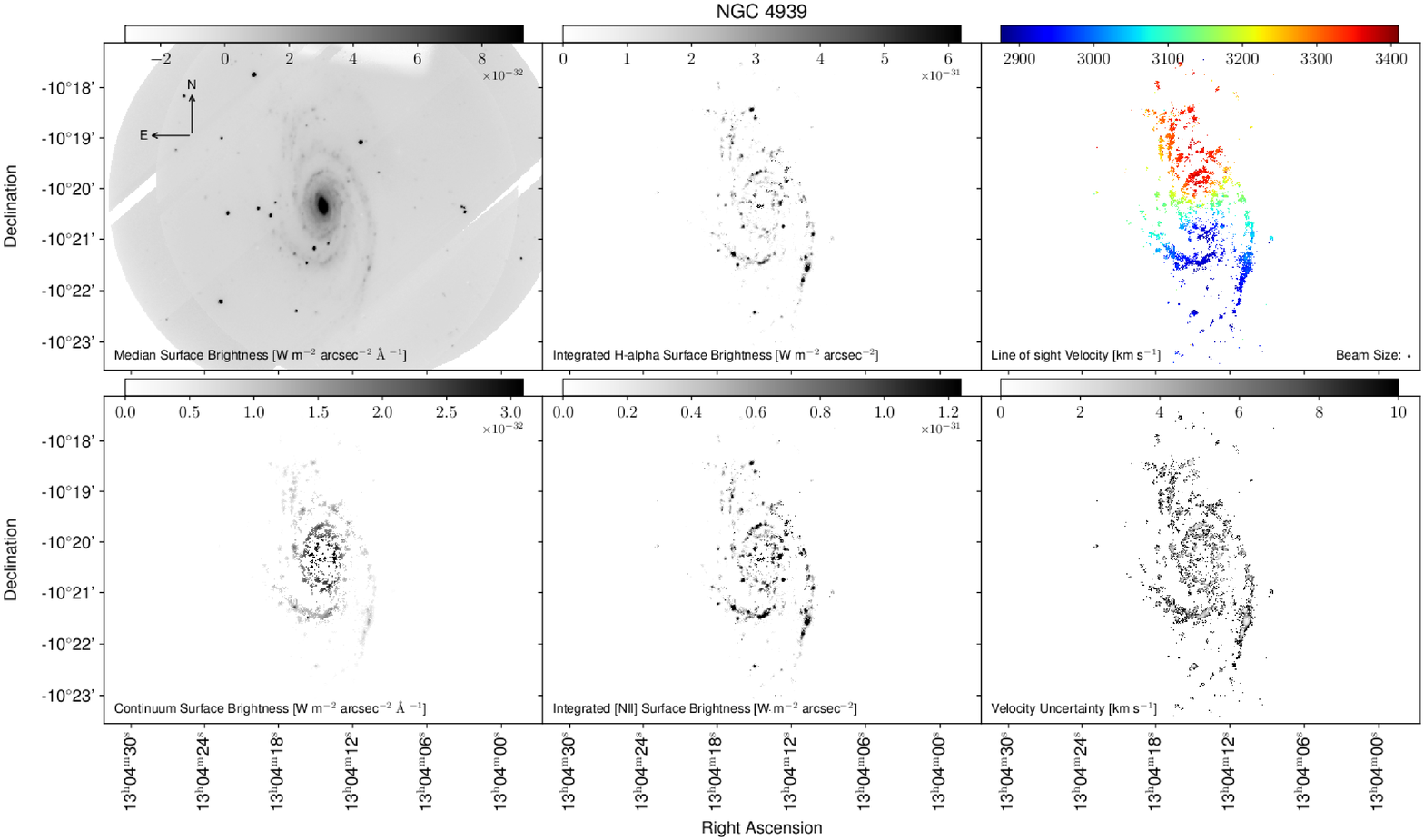}
    \includegraphics[width=\hsize]{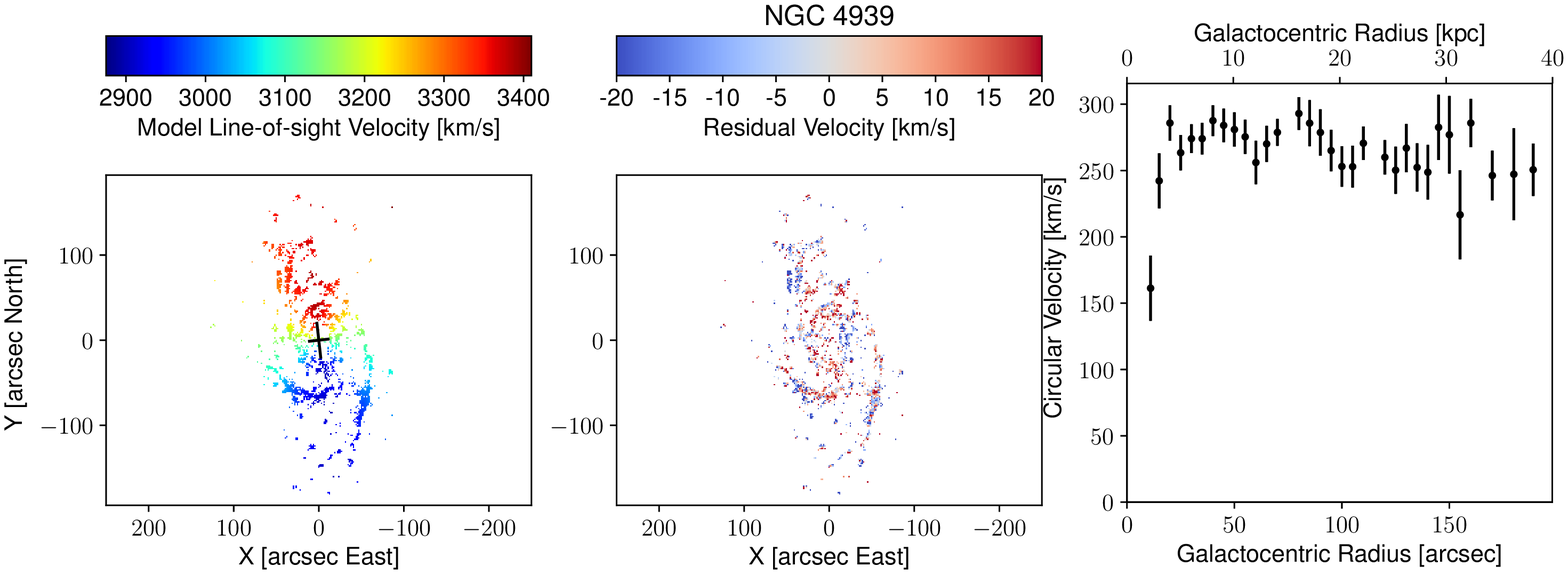}
  \end{center}
  \caption{Same as Figure \ref{fig:N337A}, but for NGC 4939.  At a distance of 41.6 Mpc, the physical scale is $202 \textrm{ pc}/\arcsec$.
    \label{fig:N4939}}
\end{figure*}

\vfill\eject

\subsection{NGC 4939}
Figure~\ref{fig:N4939} presents our results for NGC 4939, which is the
most luminous galaxy in our sample.  The rotation curve rises steeply
before becoming approximately flat for $R\ga 25\arcsec$ at a value of
270 km~s$^{-1}$ out to nearly 40 kpc in the disk plane.  Our kinematic
projection parameters and center for this galaxy agree very well (see
Figure~\ref{fig:projcomp}) with those derived from the {\it I}-band image by
\citet{RINGSPhot}.

\begin{figure*}
  \begin{center}
    \includegraphics[width=0.9\hsize]{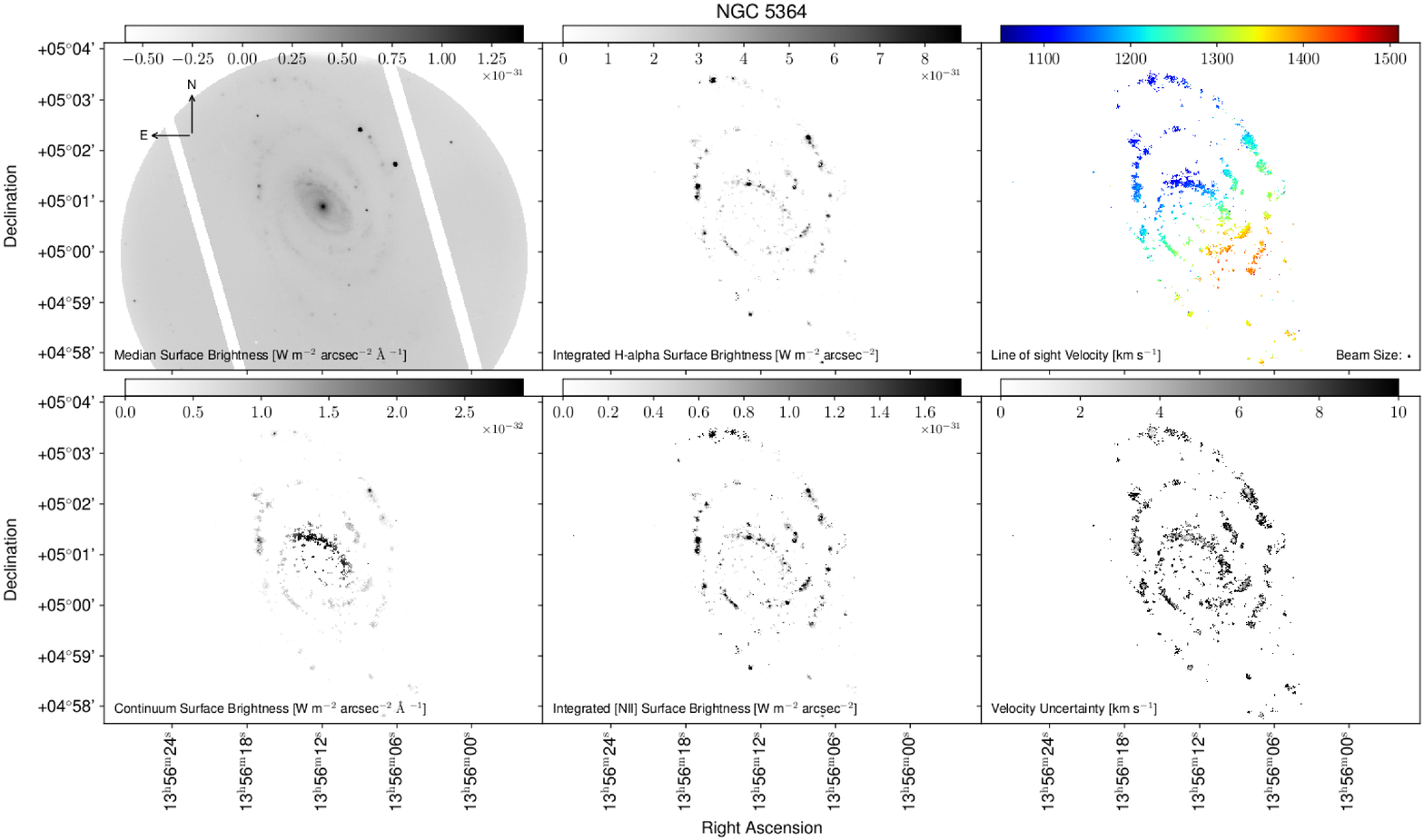} 
    \includegraphics[width=\hsize]{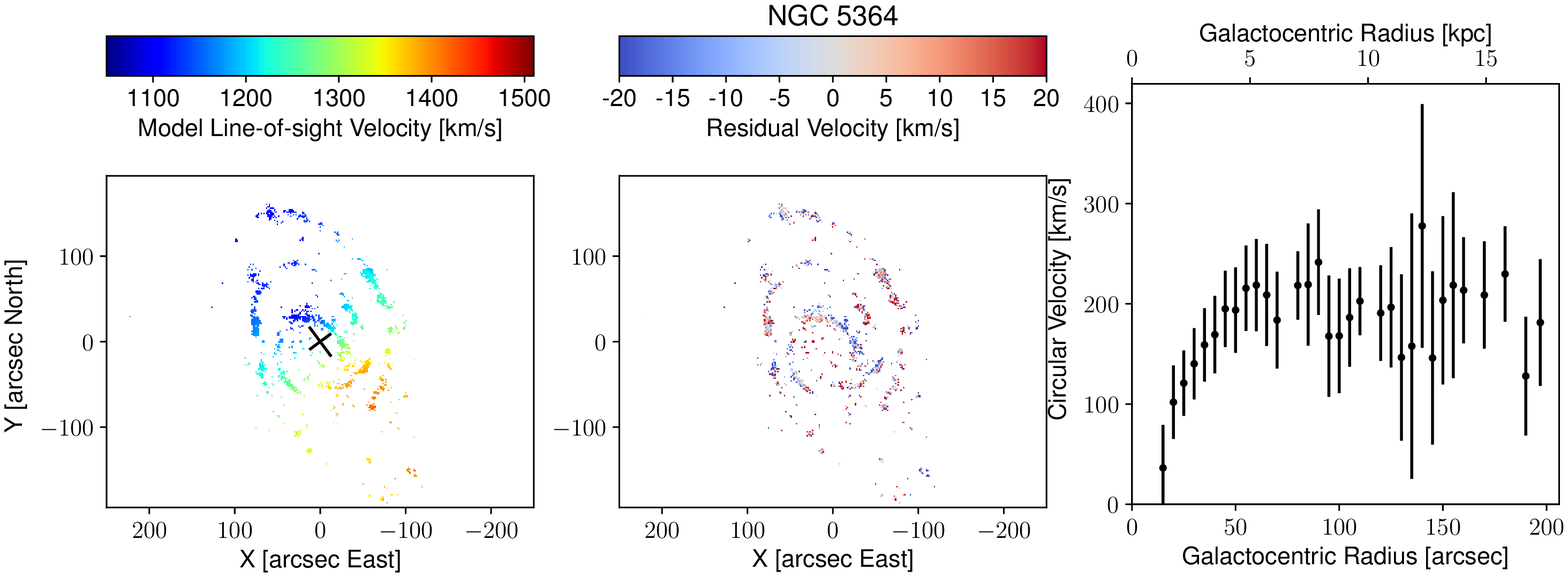}
 \end{center}
  \caption{Same as Figure \ref{fig:N337A}, but for NGC 5364.  At a
    distance of 18.1 Mpc, the physical scale is $87.8 \textrm{
      pc}/\arcsec$.
    \label{fig:N5364}}
\end{figure*}

\vfill\eject\phantom{blah}\vfill\eject

\subsection{NGC 5364}
The \Ha\ emission in NGC 5364 very strongly traces its spiral arms and
we detect no \Ha\ emission within the innermost $\sim15\arcsec$.  The
rotation curve is rising roughly linearly outside this radius before
becoming approximately flat for $R \ga 80\arcsec$.  Because the
kinematic data are somewhat sparse, the galaxy's inclination has a
moderately large uncertainty, leading to a large uncertainty in the
overall normalization of the rotation curve.

Our fitted position angle and inclination differ (see
Figure~\ref{fig:projcomp}) by a few degrees from the values derived
from the {\it I}-band image by \citet{RINGSPhot}, although differences are
not large compared with the uncertainties.  Again the lack of
kinematic data in the inner part of map led to larger than usual
uncertainties in the position of the center.

\begin{figure*}
  \begin{center}
    \includegraphics[width=0.9\hsize]{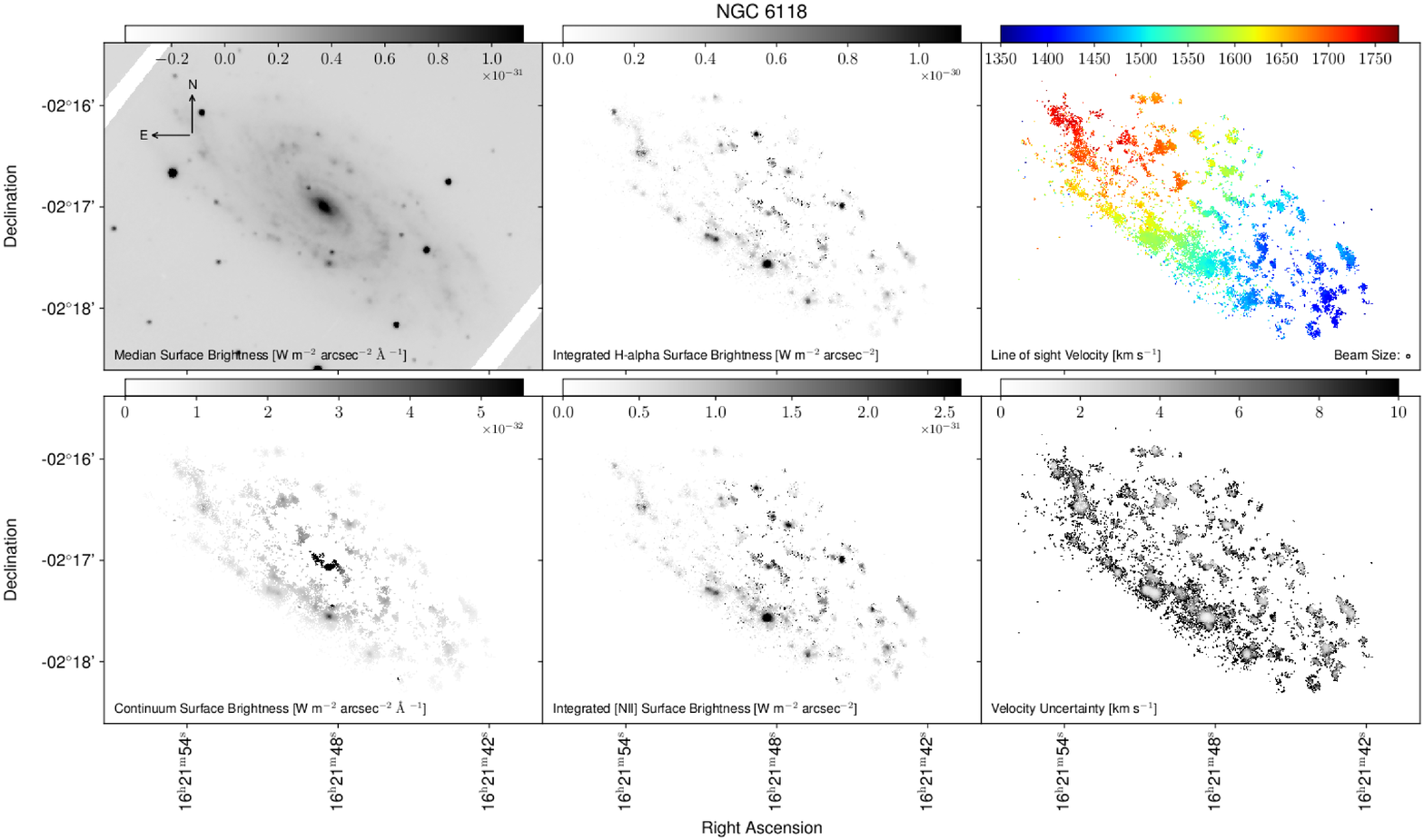}
    \includegraphics[width=\hsize]{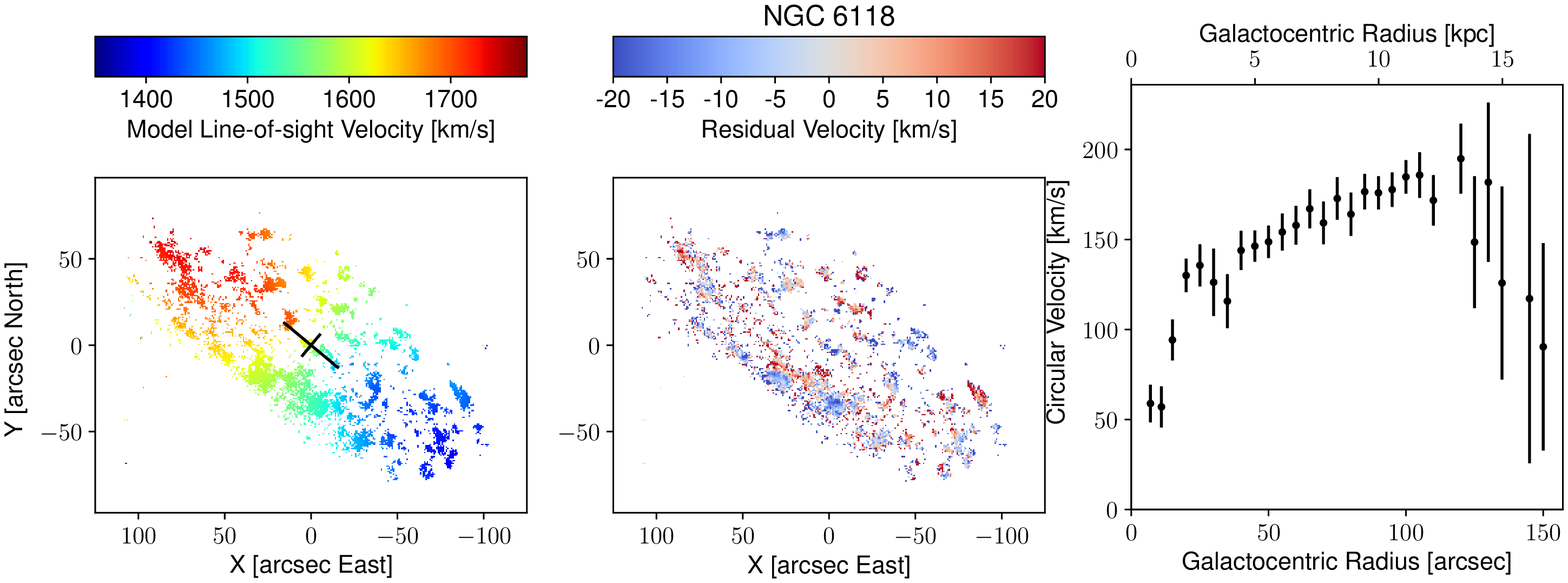}
  \end{center}
  \caption{Same as Figure \ref{fig:N337A}, but for NGC 6118.  At a
    distance of 22.9 Mpc, the physical scale is $111 \textrm{
      pc}/\arcsec$.
    \label{fig:N6118}}
\end{figure*}

\vfill\eject\phantom{blah}\vfill\eject

\subsection{NGC 6118}
Our velocity map for NGC 6118 is presented in Figure~\ref{fig:N6118}.
The rotation curve extracted from our axisymmetric model rises
continuously from the center to $R \ga 100\arcsec$.  The decreasing
values beyond this radius have large uncertainties.

Our best-fitting projection angles agree (Figure~\ref{fig:projcomp})
with the values derived from the {\it I}-band image by \citet{RINGSPhot},
but the centers disagree by about 5\arcsec, or about $6\sigma$.

Our rotation curve also agrees very well (Figure~\ref{fig:rc_comp})
with that obtained by \citet{1984A&AS...58..351M} using a longslit and
who adopted an inclination of 62\arcdeg.

\begin{figure*}
  \begin{center}
    \includegraphics[width=0.9\hsize]{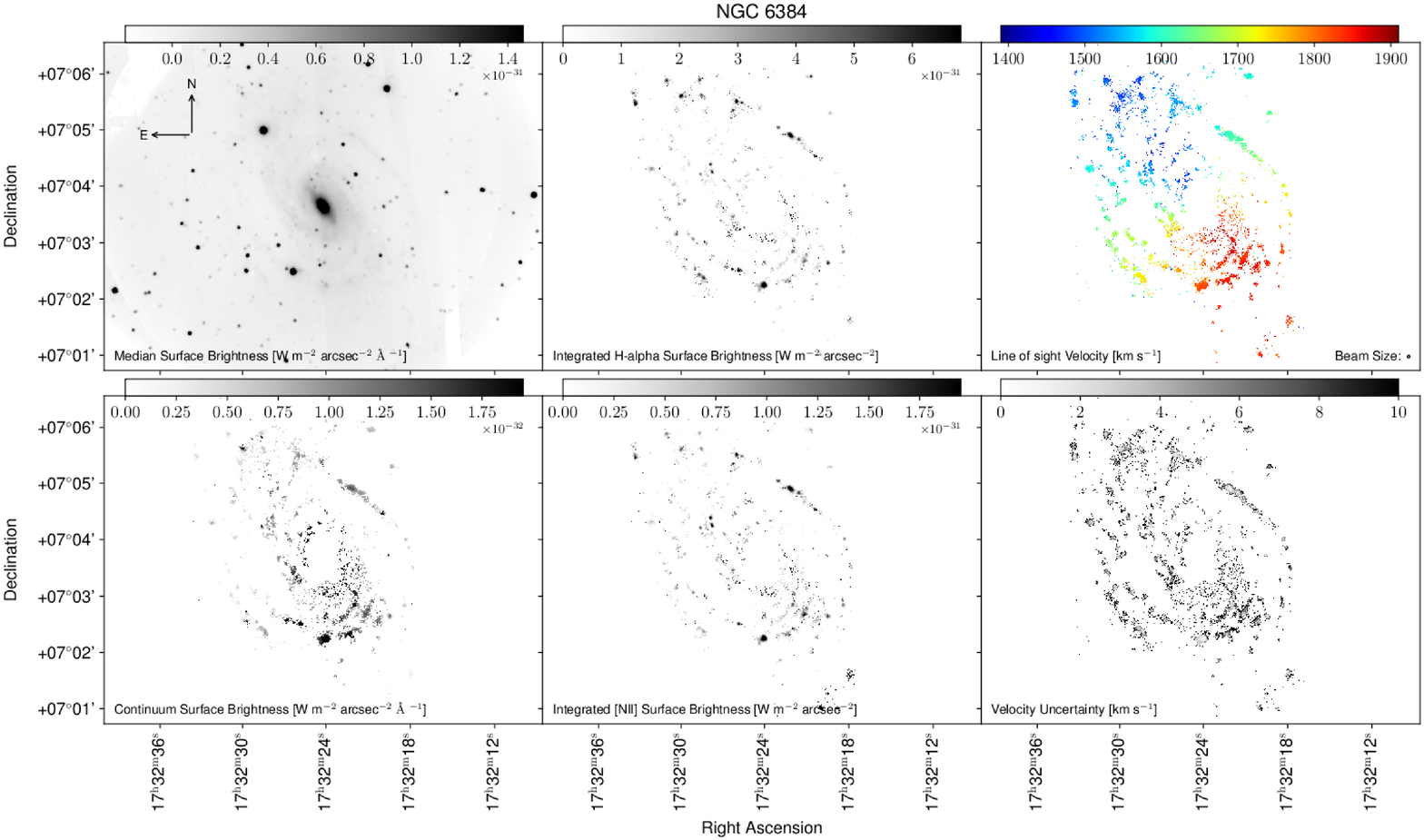}
    \includegraphics[width=\hsize]{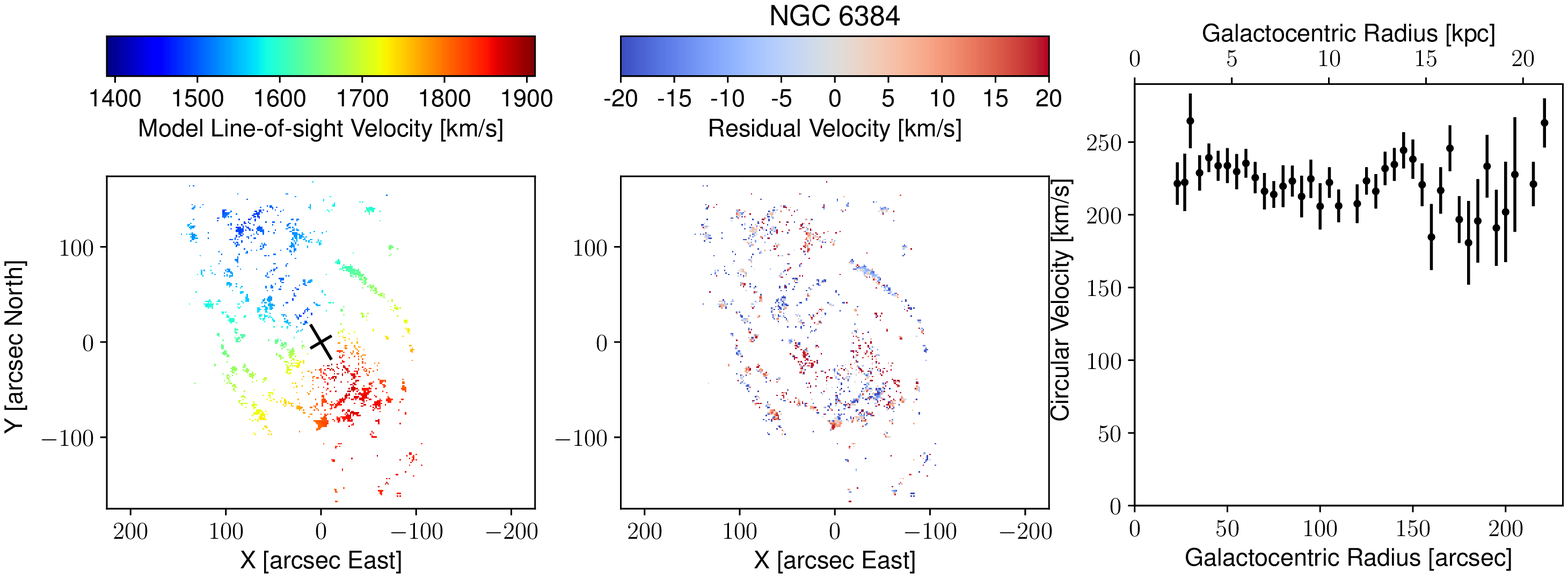}
  \end{center}
  \caption{Same as Figure \ref{fig:N337A}, but for NGC 6384.  At a
    distance of 19.7 Mpc, the physical scale is $95.5 \textrm{
      pc}/\arcsec$.
    \label{fig:N6384}}
\end{figure*}

\vfill\eject\phantom{blah}\vfill\eject

\subsection{NGC 6384}
We present our velocity map for NGC 6384 in Figure~\ref{fig:N6384}.
As in NGC 5364, the \Ha\ emission closely traces the spiral arms.  We
detect no \Ha\ emission within the innermost $\sim25\arcsec$.  Our
fitted rotation curve is roughly flat from this point to the outermost
limits of our data.

Our best-fitting model's inclination is in reasonable agreement with
the uncertain value (see Figure~\ref{fig:projcomp}) derived from the
{\it I}-band image by \citet{RINGSPhot}, while the position angle and
center are in better agreement.

Figure~\ref{fig:rc_comp} shows that our estimates of the circular
speed in NGC~6384 are systematically higher than those of
\citet{1995A&AS..110..279S}, as was the case for NGC~2280.  Again
these authors appear not have corrected their orbital speeds for
inclination.

\begin{figure*}
  \begin{center}
    \includegraphics[width=0.9\hsize]{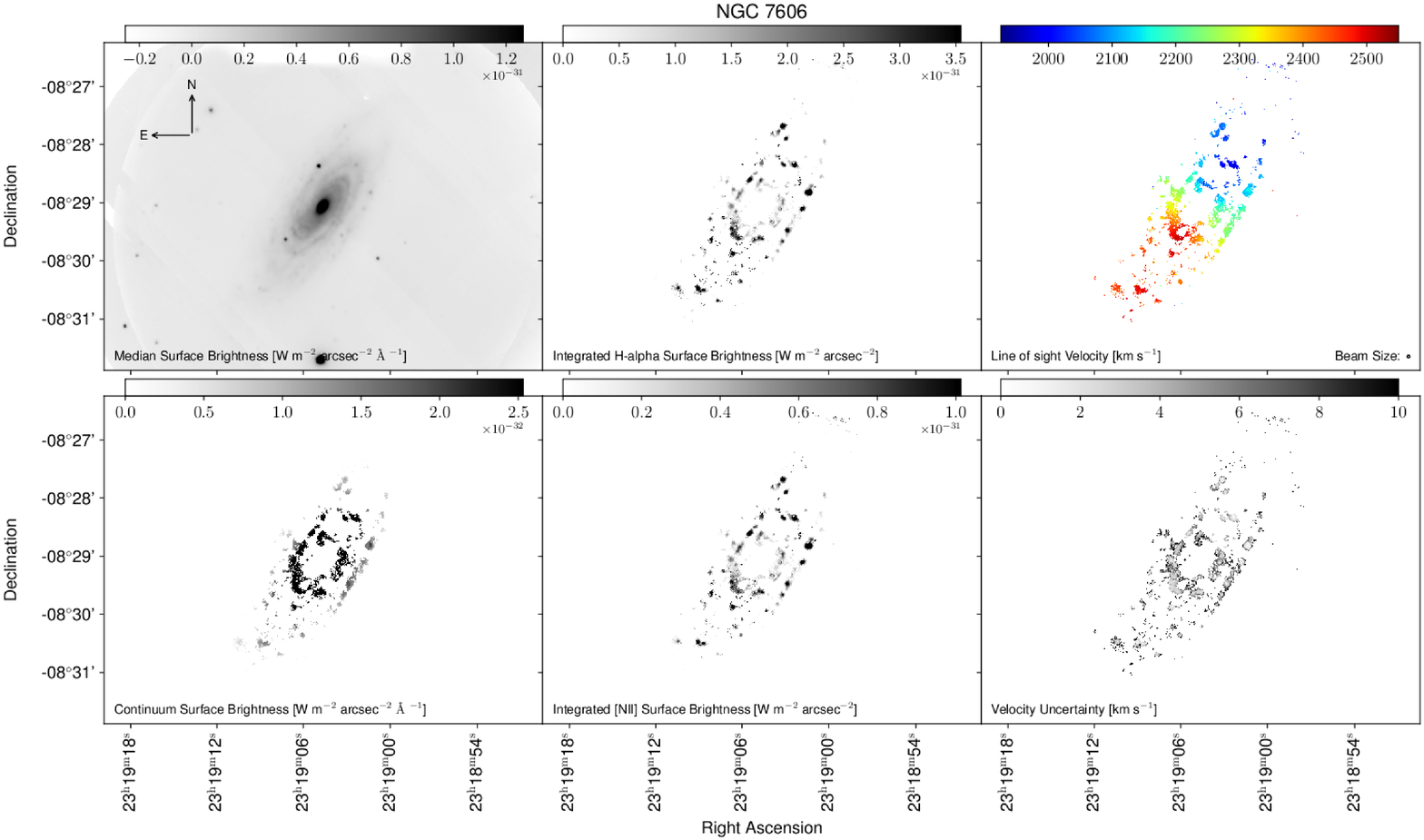}
    \includegraphics[width=\hsize]{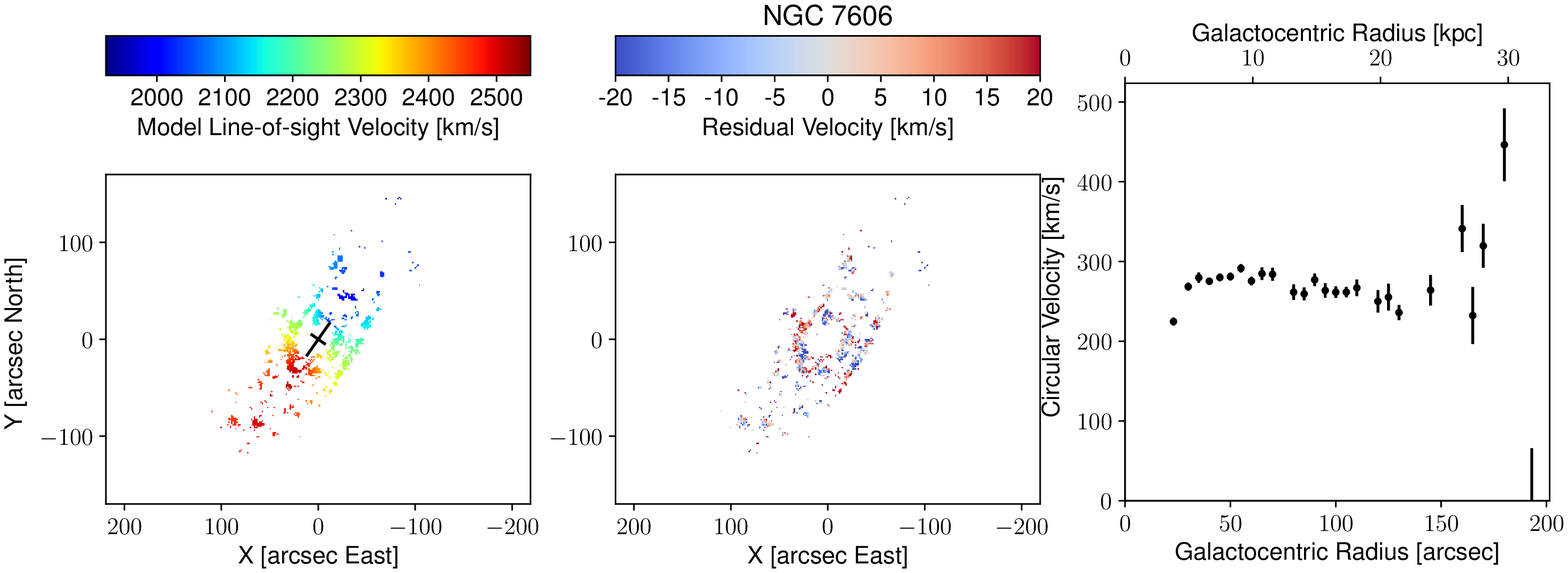}
  \end{center}
  \caption{Same as Figure \ref{fig:N337A}, but for NGC 7606.  At a
    distance of 34.0 Mpc, the physical scale is $165 \textrm{
      pc}/\arcsec$.
    \label{fig:N7606}}
\end{figure*}

\vfill\eject\phantom{blah}\vfill\eject

\subsection{NGC 7606}
NGC 7606 is the fastest-rotating galaxy in this sample and the second
most-luminous.  Again the velocity map, Figure~\ref{fig:N7606},
displays the flow pattern of a typical spiral disk, and again we
detect no \Ha\ emission in the innermost $\sim15\arcsec$.  Our fitted
rotation curve appears to be rising from our innermost point, becoming
roughly flat from $R\sim30\arcsec$, before declining somewhat from
$\sim50$--$120\arcsec$ with a hint of an outer increase, although the
uncertainties are large due to the sparseness of our data at these
radii.

The inclination and position angle of this galaxy are extremely
tightly constrained by our kinematic models and agree very well,
Figure~\ref{fig:projcomp}, with the projection angles derived from the
{\it I}-band image by \citet{RINGSPhot}, as does the location of the center
despite the absence of data at small radii.

In general, our rotation curve measurements agree well with the
previous measurements by \citet{1982ApJ...261..439R} (blue points in
Figure~\ref{fig:rc_comp}) and \citet{1996ApJS..107...97M} (red
points), who adopted inclinations of 66\arcdeg\ and
70\arcdeg\ respectively.

\begin{figure*}
  \begin{center}
    \includegraphics[width=0.9\hsize]{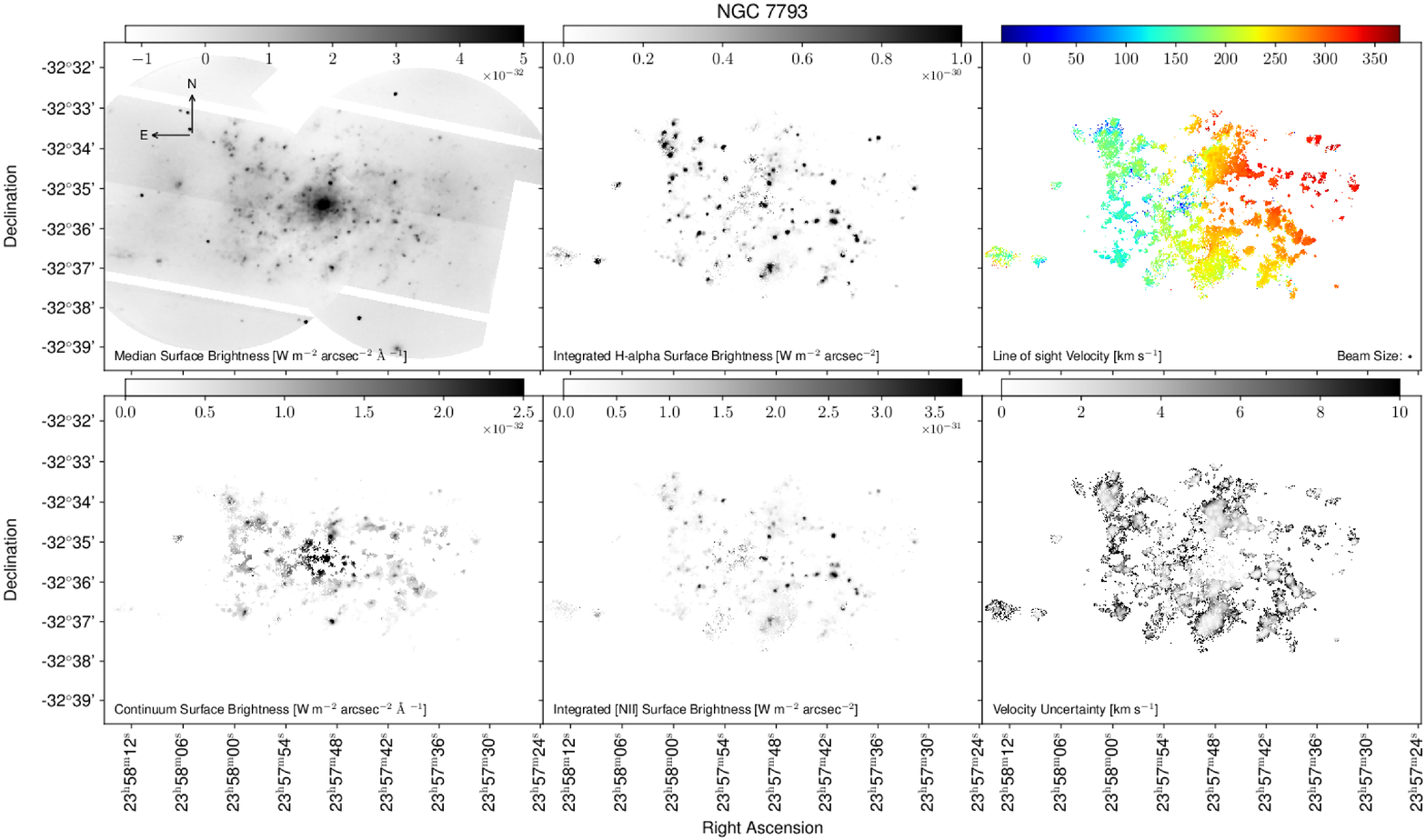}
    \includegraphics[width=\hsize]{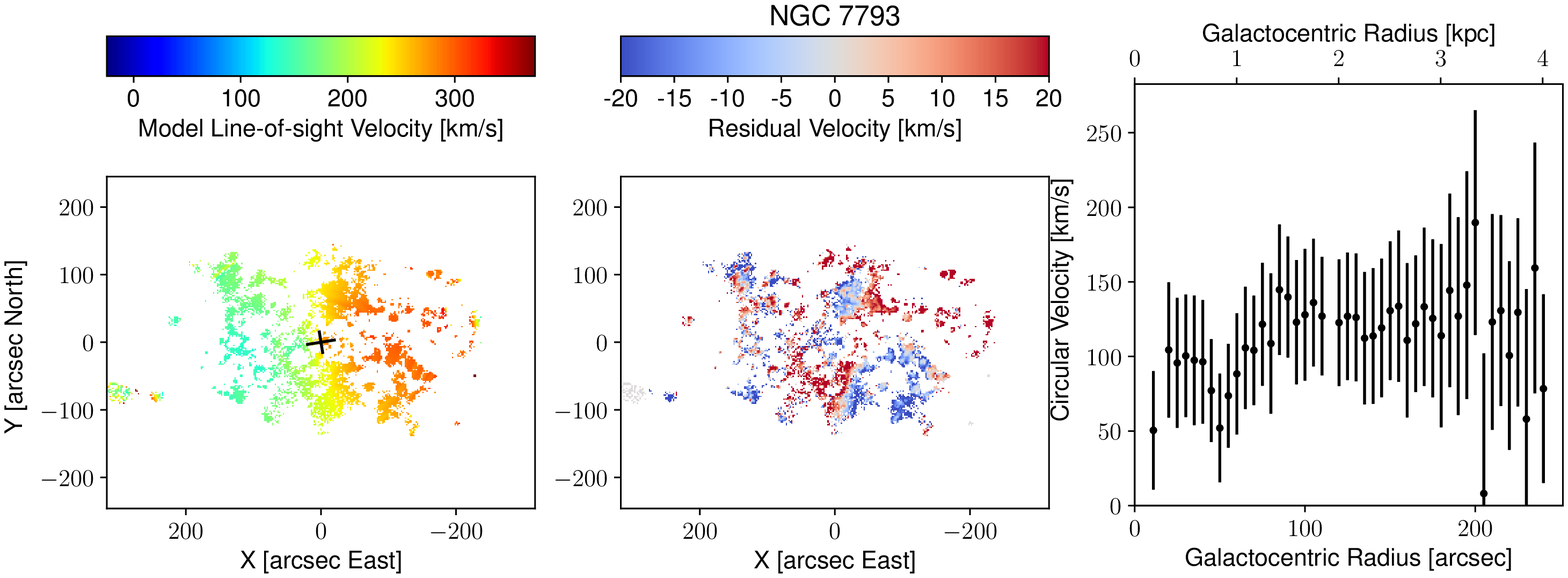}
  \end{center}
  \caption{Same as Figure \ref{fig:N337A}, but for NGC 7793.  At a
    distance of 3.44 Mpc, the physical scale is $16.7 \textrm{
      pc}/\arcsec$.  Because we obtained 4 observations of the East
    (approaching) side of this galaxy and only 1 observation of the
    West (receding) side, our sensitivity is significantly stronger on
    the Eastern portion of these maps.  All 5 observations overlap in
    the central region.
    \label{fig:N7793}}
\end{figure*}

\vfill\eject

\subsection{NGC 7793}
NGC 7793 has the largest angular size of our sample and the velocity
map, Figure~\ref{fig:N7793}, was derived from the combination of two
separate pointings.

Our fitted rotation curve shows a general rise to $R\sim 100\arcsec$,
except for a slight decrease around $R\sim40\arcsec$.  Our data in the
outermost parts of the galaxy are too sparse to measure the orbital
speed reliably.  As for NGC 337A, the large uncertainties on
individual points in the rotation curve are mostly due to the large
uncertainty in NGC 7793's inclination in our model.

The projection parameters of our best-fitting model agree well within
the larger than usual uncertainties, Figure~\ref{fig:projcomp}, with
those derived from the {\it I}-band image by \citet{RINGSPhot}, and while
our fitted center is some 14\arcsec\ from the photometric center, our
uncertainty estimates are also large, so that this discrepancy is
$2.5\sigma$.

\begin{figure*}[t]
  \begin{center}
    \includegraphics[width=0.69\hsize]{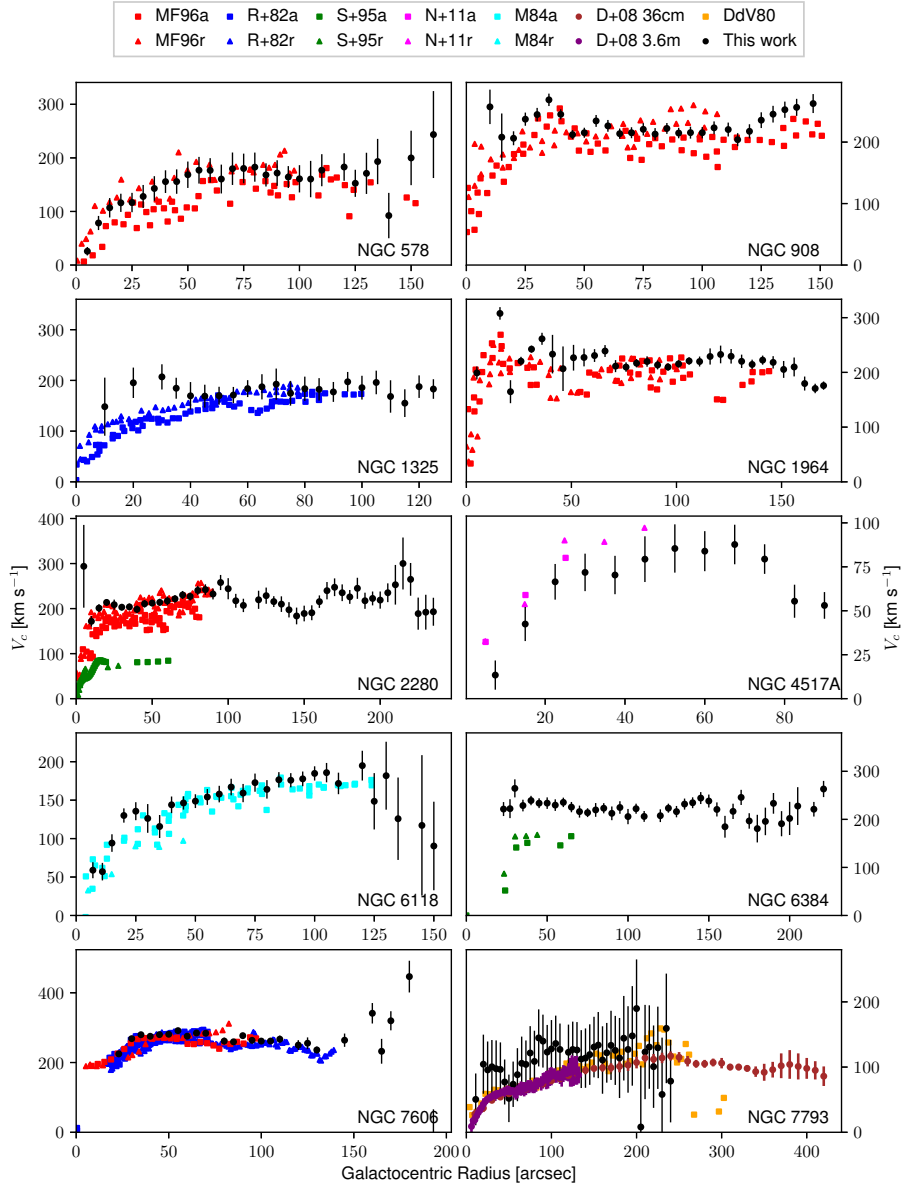}
  \end{center}
  \caption{A comparison of our best-fitting model rotation curves
    (black circles with error bars) to previous measurements from the
    literature.  In all cases, squares are from the approaching side of
    the galaxy, triangles from the receding side, and circles from an
    azimuthal average.  Unless otherwise specified, we have used own
    own best-fitting values for systemic velocity and inclination (see
    Table \ref{tab:tab2}) to deproject the data.  Red points (NGC 578,
    908, 1964, 2280, and 7606): \citet{1996ApJS..107...97M} via \Ha\
    longslit spectroscopy (Note: We have adopted a systemic velocity
    of 1960 km~s$^{-1}$ for NGC 1964 rather than our best-fitting
    value to match the authors' spectra.  The authors also report a
    rotation curve for NGC 1325, but the wavelength calibration for
    those data appears to have been incorrect.).  Blue points (NGC 1325
    and 7606): \citet{1982ApJ...261..439R} via \Ha\ and \N2 longslit
    spectroscopy.  Green points (NGC 2280 and 6384):
    \citet{1995A&AS..110..279S} via \Ha\ and \N2 longslit
    spectroscopy.  Magenta points (NGC 4517A):
    \citet{2011MNRAS.413.1875N} via \Ha\ IFU spectroscopy.  Cyan points
    (NGC 6118): \citet{1984A&AS...58..351M} via optical longslit
    spectroscopy.  Brown and purple points (NGC 7793):
    \citet{2008AJ....135.2038D} via \Ha\ \FP\ spectrophotometry.  Orange
    points (NGC 7793): \citet{1980ApJ...242...30D} via \Ha\ \FP\
    spectrophotometry.
    \label{fig:rc_comp}}
\end{figure*}

Again in Figure~\ref{fig:rc_comp} we compare our estimated rotation
curve with those previously reported by \citet{1980ApJ...242...30D}
(orange points) and by \citet{2008AJ....135.2038D} (purple points),
who adopted inclinations of 53\arcdeg\ and 46\arcdeg\ respectively
that are both larger than our 40\arcdeg.  Consequently, our estimated
speeds are above theirs at most radii.  The shapes of the rotation
curves are generally similar, although we find a steeper inner rise.

\subsection{Discussion}
As we have discussed for the individual cases, the rotation curves we
derive from fitting axisymmetric flow patterns to our velocity maps
agree quite well with previously published estimates from several
different authors and using a number of different optical instruments.
These comparisons are shown in Figure~\ref{fig:rc_comp}, where most
systematic discrepancies can be attributed to differences between the
inclinations we adopt, and those in the comparison work.  This
generally good agreement is reassuring.

\subsection{Oval disks?}
Discrepancies between the position angle and inclination fitted
separately to a kinematic map and a photometric image of the same
galaxy would be expected if the disk were intrinsically oval, as has
been claimed in some cases \citep[e.g][]{2011ApJ...739L..27P} and
emphasized as a possibility by \citet{2013seg..book....1K}.  Even were
the projected major axis to be closely aligned with either of the
principal axes of a strongly oval disk, the fitted inclinations should
differ.

We have no clear evidence of this behavior in our sample of galaxies,
since the projection angles derived from fitting axisymmetric models
to our velocity maps generally agree, within the estimated
uncertainties, with those fitted to the {\it I}-band images
\citep{RINGSPhot}, as shown in Figure~\ref{fig:projcomp}.  We argued
above that the discrepancy in NGC~337A is due to the faintness of the
outer disk, while those in NGC~578 and NGC~908 can be ascribed to
asymmetries.  Note that \citet{2003AJ....125.1164B} reported that the
position angle of the galaxy major axis estimated from photometric
images and kinematic maps never exceeded 4\arcdeg\ in their larger
sample of 74 galaxies, and \citet{2014A&A...568A..70B} found only
minor misalignments in a sample of intrinsically barred galaxies.
Since these were all randomly selected spiral galaxies, it would seem
that the incidence of intrinsically oval disks is low, at least over
the radial extent of these maps.

Futhermore, \citet{2003AJ....125.1164B}, found that the kinematic
centers of their models were within 2\farcs7 of the photmetric
centers in 67 out of 74 galaxies in their sample.  Here we find the
centers of our kinematic models are consistent in several cases with
the photometric centers (see Figure~\ref{fig:projcomp}), and the
greater discrepancies generally arise where our maps are sparse or
lack data in the center.

\begin{figure}
  \begin{center}
    \bigskip
    \includegraphics[width=\hsize]{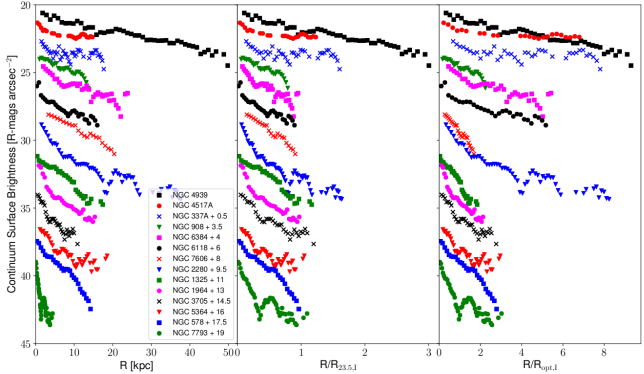}
  \end{center}
  \caption{Left: Azimuthally averaged \textit{R}-band continuum
    surface brightness profiles plotted as functions of galactocentric
    radius in kpc.  Center: The same values plotted as functions of
    galactocentric radius rescaled by each galaxy's $R_{23.5}$ in the
    \textit{I}-band.  Right: The same values plotted as functions of
    galactocentric radius rescaled by each galaxy's $R_{\rm opt}$ in the
    \textit{I}-band.  In each panel, the lines have been vertically
    offset by a constant to separate them.
    \label{fig:sb}}
\end{figure}

\section{Radial trends}
Figure \ref{fig:sb} shows the azimuthally averaged \textit{R}-band
continuum surface brightness of our galaxies derived from our
\Ha\ \FP\ data cubes plotted against three different measures of
galactocentric radius.  These surface brightness profiles assume that
the disk projection parameters are those of the best-fitting
\textit{I}-band models of \citet{RINGSPhot}.  The surface brightness
profiles show qualitative and quantitative agreement with the
\textit{R}-band surface brightness profiles of \citet{RINGSPhot}, but
have a smaller radial extent.

\begin{figure}
  \begin{center}
    \bigskip
    \includegraphics[width=\hsize]{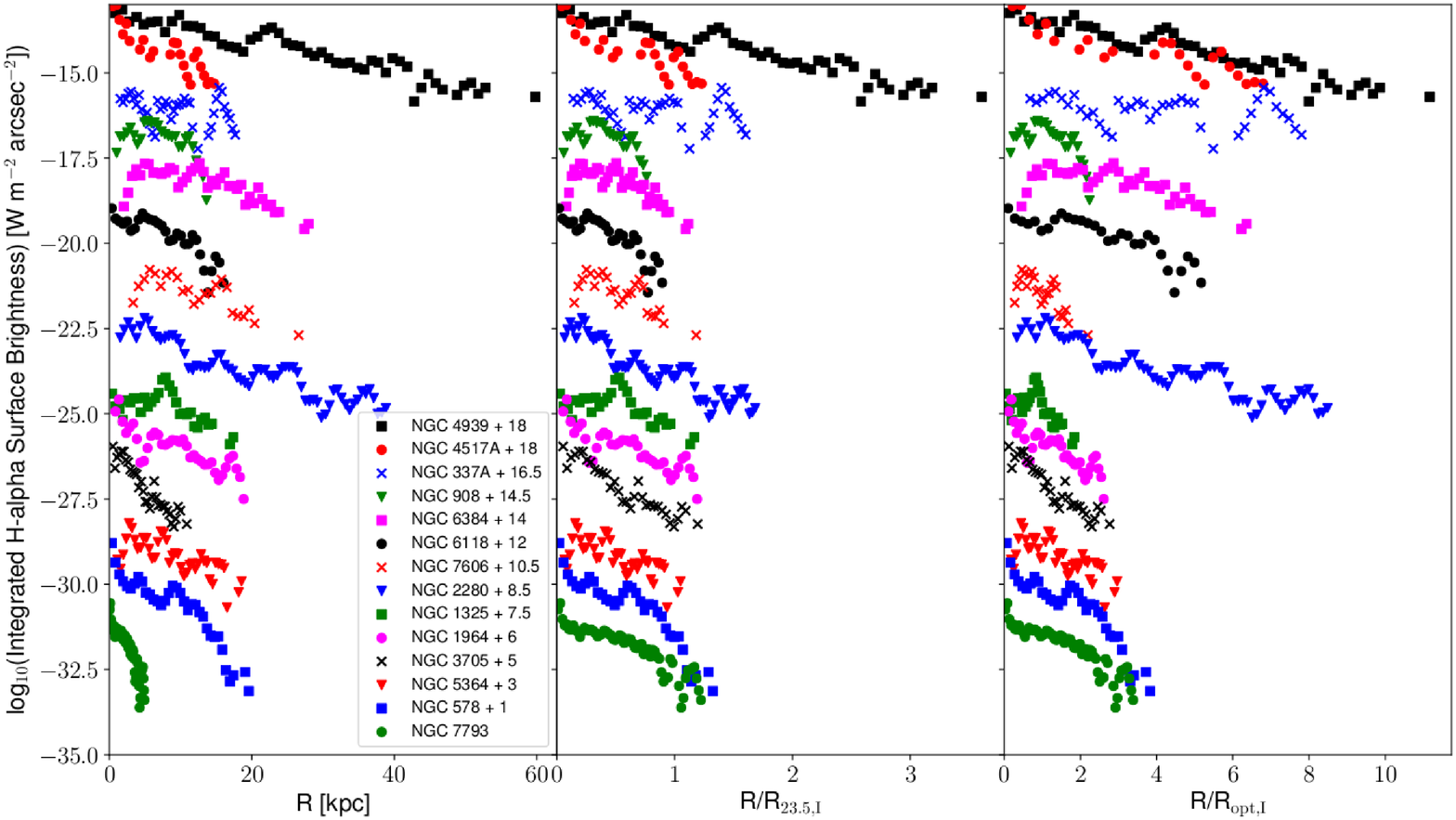}
  \end{center}
  \caption{Left: Azimuthally averaged integrated \Ha\ surface
    brightness profiles plotted as functions of galactocentric radius
    in kpc.  Center: The same values plotted as functions of
    galactocentric radius rescaled by each galaxy's $R_{23.5}$ in the
    \textit{I}-band.  Right: The same values plotted as functions of
    galactocentric radius rescaled by each galaxy's $R_{\rm opt}$ in the
    \textit{I}-band.  In each panel, the lines have been vertically
    offset by a constant to separate them.
    \label{fig:haprofiles}}
\end{figure}

Figure \ref{fig:haprofiles} shows the azimuthally averaged integrated
\Ha\ surface brightnesses of our galaxies, i.e.\ the values of $F_H$
in Equation \ref{eqn:model}.  These values should be considered as
lower limits on the true \Ha\ intensity, as the averages were taken
over all pixels in a radial bin, including those which fell below our
signal-to-noise threshold.

\begin{figure}
  \begin{center}
    \medskip
    \includegraphics[width=\hsize]{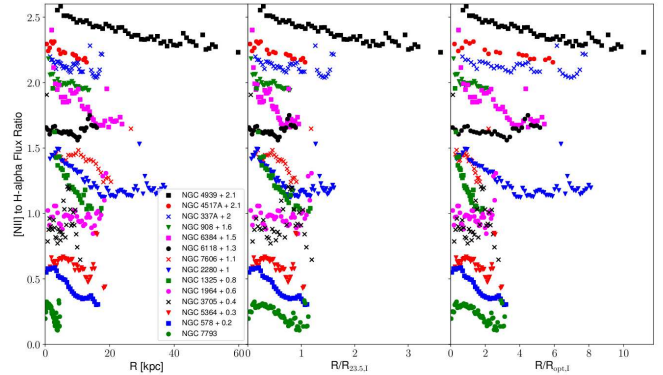}
  \end{center}
  \caption{Left: Azimuthally averaged N2 Index ($\mathrm{N2}\equiv
    \log(\mathrm{F_{\mathrm{N2~6583}}/\mathrm{F_{H\alpha}}})$) plotted
    as functions of galactocentric radius in kpc.  Center: The same
    values plotted as functions of galactocentric radius rescaled by
    each galaxy's $R_{23.5}$ in the \textit{I}-band.  Right: The same
    values plotted as functions of galactocentric radius rescaled by
    each galaxy's $R_{\rm opt}$ in the \textit{I}-band.  In each panel,
    the lines have been vertically offset by a constant to separate
    them.
    \label{fig:n2ratios}}
\end{figure}

\vfill\eject
\subsection{\N2-to-\Ha\ Ratio and Oxygen Abundance}
Figure \ref{fig:n2ratios} shows the azimuthally averaged value of the
ratio of the integrated \N2 6583 surface brightness to the integrated
\Ha\ surface brightness, commonly known as the ``N2 Index''
\citep{1979A&A....78..200A}
\begin{equation}
\mathrm{N2} \equiv
\log(F_{\mathrm{N2}~6583}/F_{\mathrm{H\alpha}}).
\end{equation}
It is important to note that the plotted quantity is the average value
of the ratio ($ \langle F_N/F_H \rangle $) and not the ratio of the
averages ($\langle F_N \rangle / \langle F_H \rangle $).  We note that
all of our galaxies show a downward trend in this parameter.  The
relative intensities of these two lines are complicated functions of
metallicity and electron temperature in the emitting gas, and the line
intensity ratio is also known to be sensitive to the degree of
ionization of the gas \citep{1983MNRAS.204...53S}.

\begin{figure}
  \begin{center}
    \bigskip
    \includegraphics[width=.96\hsize]{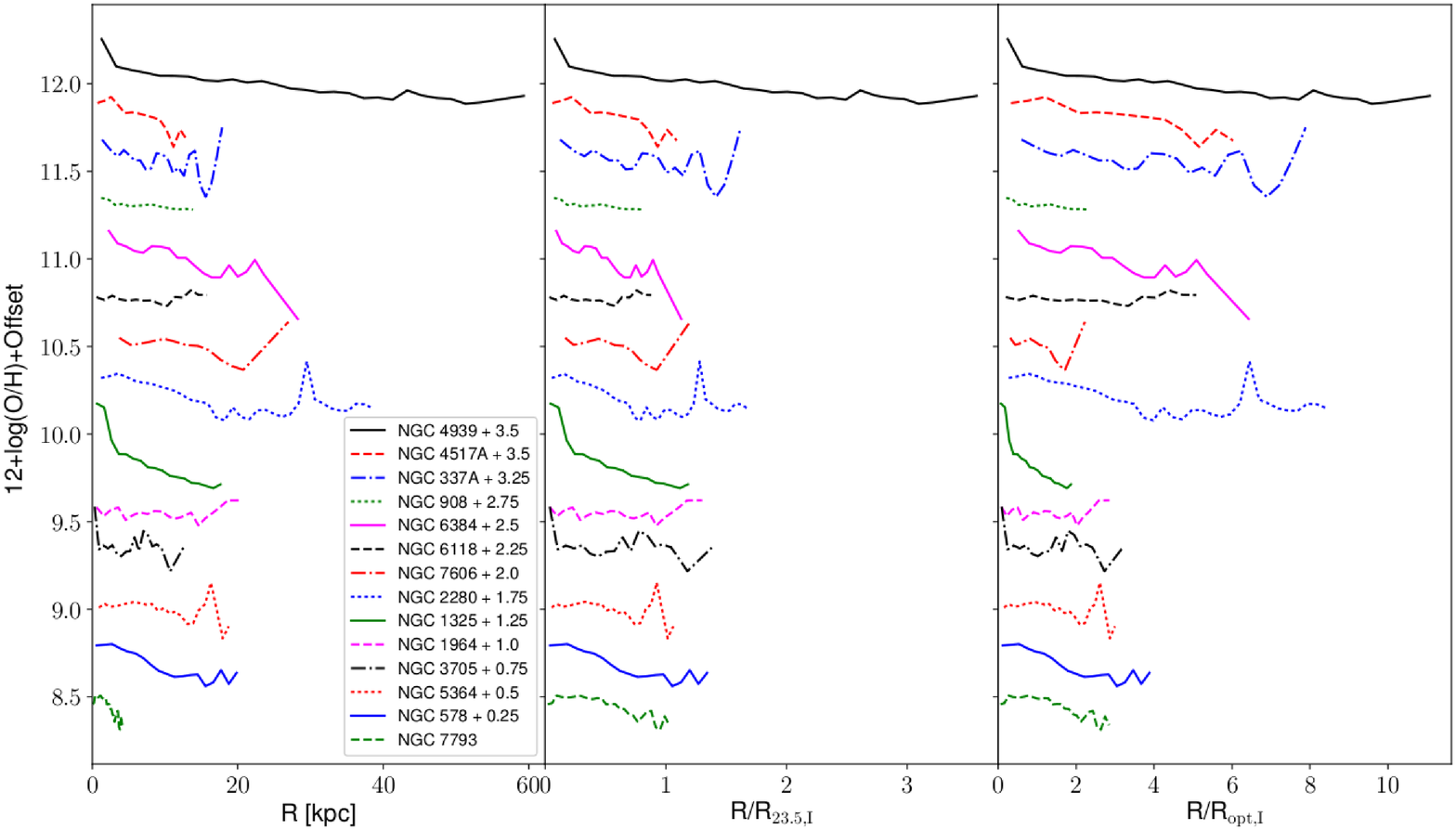}
  \end{center}
  \caption{Left: Azimuthally averaged oxygen abundances
    ($12+\log(\mathrm{O}/\mathrm{H})$) plotted as functions of
    galactocentric radius in kpc.  Center: The same values plotted as
    functions of galactocentric radius rescaled by each galaxy's
    $R_{23.5}$ in the \textit{I}-band.  Right: The same values plotted
    as functions of galactocentric radius rescaled by each galaxy's
    $R_{\rm opt}$ in the \textit{I}-band.  In each panel, the lines have
    been vertically offset by a constant to separate them.
    \label{fig:O2}}
\end{figure}

Because this ratio is sensitive to the metallicity of a galaxy and
does not strongly depend on absorption, it has been widely used as an
indicator of oxygen abundance \citep[e.g.][]{2009MNRAS.398..949P,
  2013A&A...559A.114M}; \citet{2004MNRAS.348L..59P} show that the data
support an approximately linear relation between oxygen abundance and
N2 index, which holds over the range $-2 \ga \mathrm{N2} \ga -0.5$,
but the relation may steepen at both higher and lower values of the
ratio.  \citet{2013A&A...559A.114M} give the following relation
between the N2 index and oxygen abundance:
\begin{equation}
   12+\log(\mathrm{O}/\mathrm{H}) = 8.743 + 0.462\times\mathrm{N2}.
\label{eq.oabund}
\end{equation}
We have used this relation to derive the mean radial variation of
oxygen abundance in our galaxies displayed in Figure \ref{fig:O2}. As
in many previous studies, we find that our galaxies generally manifest
a declining trend in metallicity \citep[e.g.][]{1992MNRAS.259..121V,
  1994ApJ...420...87Z, 2010ApJS..190..233M, 2017MNRAS.469..151B}.
With the exception of NGC~6384, the most extended normalized profiles
(e.g. NGC~4939, NGC~337A, NGC~4939, NGC~2280) show hints of a
flattening in the outer parts, as has also been reported for large
samples \citep[e.g.][]{2014A&A...563A..49S, 2016A&A...587A..70S}.  NGC
4939 is the only galaxy discussed in this work known to host an active
galactic nucleus (AGN).  Away from the nucleus of this galaxy, and in
all other galaxies in our sample, most ionizing radiation probably
comes from hot, young stars.  The extra ionizing radiation from the
AGN in NGC 4939 may be the reason for central spike in the apparent
oxygen abundance in this case.

\section{Summary}
\label{sec:summary}
We have presented high spatial resolution ($\sim$2.5\arcsec)
\Ha\ velocity fields of 14 of the 19 galaxies in the RINGS sample, as
well as maps of these galaxies' \textit{R}-band continuum emission and
\Ha\ and \N2 integrated surface brightness.  Additionally, we have
presented azimuthally averaged integrated surface brightness profiles
of these emission lines.  We observe a general downward trend of the
\N2-to-\Ha\ emission ratio with radius in all of our galaxies.

We have used the \df\ software package of \citet{2007ApJ...664..204S}
and \citet{2010MNRAS.404.1733S} to model the velocity fields presented
in this work.  From these models, we have extracted rotation curves at
high spatial resolution and have shown good general agreement with
those previously published, where available.  In most cases, the
projection geometries of these models agree well with the photometric
models of \citet{RINGSPhot}.  This agreement argues against the disks
being intrinsically oval.

As of 2015 Sept, the medium-resolution \FP\ etalon of SALT RSS is no
longer available for observations due to deterioration of the
reflective coatings.  The remaining five galaxies of the RINGS sample
are scheduled to be completed in the \FP\ system's high-resolution
mode.

\acknowledgments 
We thank Tad Pryor for several productive conversations in designing
our data reduction procedures, and an anonymous referee for a thorough
and constructive report.  This work was supported by NSF grant
AST/12117937 to JAS \& TBW and NSF grant PHY/1263280.  This research
made use of the NASA/IPAC Extragalactic Database (NED) which is
operated by the Jet Propulsion Laboratory, California Institute of
Technology, under contract with the National Aeronautics and Space
Administration; Astropy, a community-developed core Python package for
Astronomy \citep{2013A&A...558A..33A}; matplotlib, a Python library
for publication quality graphics \citep{Hunter:2007}; SciPy
\citep{jones_scipy_2001}; IRAF, distributed by the National Optical
Astronomy Observatory, which is operated by the Association of
Universities for Research in Astronomy (AURA) under cooperative
agreement with the National Science Foundation
\citep{1993ASPC...52..173T}; and PyRAF, a product of the Space
Telescope Science Institute, which is operated by AURA for NASA.  The
observations reported in this paper were obtained with the Southern
African Large Telescope (SALT) under programs 2011-3-RU-003,
2012-1-RU-001, 2012-2-RU-001 (PI: TBW), 2013-2-RU\_RSA-001,
2014-1-RU\_RSA-001, 2014-2-SCI-012, and 2015-1-SCI-016 (PI: JAS).

\bibliographystyle{apj}
\bibliography{references}

\end{document}